\documentclass[aps,prb,twocolumn]{revtex4-1}  

\usepackage{amsmath,amssymb}
\usepackage{graphicx}
\PassOptionsToPackage{caption=false}{subfig} 
\usepackage{subfig}
\usepackage{verbatim}
\usepackage{soul} 

\begin{document}

\title{Limit-Cycles and Chaos in the Current Through a Quantum Dot}

\author{Carlos L\'opez-Mon\'is$^1$, Clive Emary$^2$, Gerold Kiesslich$^2$, Gloria Platero$^1$, and Tobias Brandes$^2$}
\affiliation{$^{1}$Instituto de Ciencia de Materiales de Madrid, CSIC, Sor Juana In\'es de la Cruz 3, Cantoblanco, 28049 Madrid, Spain \\ $^{2}$Institute f\"ur Theoretische Physik, TU Berlin, Hardenbergstr. 36, D-10623 Berlin, Germany}

\date{16 December 2011}

\begin{abstract}
We investigate non-linear magneto-transport through a single level quantum dot coupled to ferromagnetic leads, where the electron spin is coupled to a large, external (pseudo-)spin via an anisotropic exchange interaction. We find regimes where the average current through the dot displays  self-sustained oscillations that reflect the limit-cycles and chaos and map the dependence of this behaviour on magnetic field strength and the tunnel coupling to the external leads. 
\end{abstract}

\maketitle

\section{Introduction}

Single-electron transport through nanostructures has developed into a powerful spectroscopic tool for probing  correlations, quantum coherence, and interactions with the environment on a microscopic level \cite{Grabert,Hanson_RMP_2007,Shinkai_PRL_2009,Andergassen10}. Some recent examples include experiments with semiconductor quantum dots that have provided detailed insight into level structures \cite{Reed_PRL_1988,Reimann_RMP_2002,Hanson_RMP_2007}, Coulomb and spin-blockade effects \cite{Ono_Science_2002}, phonon emission \cite{Roulleau_Nature_2011}, or the statistics of individual electron tunnel events \cite{Nazarov,*Fujisawa_Science_2006,*Gustavsson_PRL_2006}. 

In this paper, we propose the time-dependent, average current of electrons through a single level quantum dot as probe for classical non-linear dynamics and chaos \cite{Strogatz}. Specifically, we consider  electronic magneto-transport through a quantum dot containing two spin-split levels with an anisotropic coupling between the electron spin and an external, classical magnetic moment or pseudo-spin. In order to have a spin-polarized current through the quantum dot, we consider ferromagnetic leads (see Ref.~[\onlinecite{Weymann_PRB_2005}] and references there in).


Previous works have analyzed the anisotropic interaction between two spins in a {\em closed} system under an external magnetic field~\cite{Feingold_PhysD_1983,Peres_PRA_1984,*Feingold_PRA_1984,Magyari_ZPhysB_1987,Robb_PRE_1998}, showing either regular (integrable) or chaotic (nonintegrable) classical orbits. 
The results presented here demonstrate that the signatures of non-linear dynamics and classical chaos of the closed system also persist in the non-equilibrium regime, where the additional coupling to the electronic reservoirs leads to an even richer dynamics that can be probed, e.g., by varying the magnetic field and the tunnel rates. In particular, one finds a transition from a regime with damped current transients and a constant current, to a situation where the current displays self-sustained regular limit-cycle oscillations, or chaotic behaviour. Limit-cycles in transport have also been found recently in theoretical calculations in mesoscopic systems coupled to mechanical degrees of freedom~\cite{Novotny_PRL_2003,Pistolesi_PRL_2005,Hussein_PRB_2010,Bode_PRL_2011}.

Experimental inspiration for our model comes from the hyperfine interaction in quantum dots.
%
%
The interaction of electron spins in quantum dots with surrounding nuclear spins is usually viewed as simply giving rise to spin relaxation and decoherence \cite{Khaetskii_PRL_2002,Reilly_Science_2008}.
%
Recently, however, transport experiments through semiconductor double quantum dots have shown non-linear current behaviour which has been attributed to hyperfine interaction inducing a dynamical nuclear spin polarization~\cite{Ono_PRL_2004,*Koppens_Science_2005,Baugh_PRL_2007}. 
The feedback between electron and nuclei spin polarization gives rise to nontrivial features in the current, including self-sustained oscillations~\cite{Ono_PRL_2004,*Koppens_Science_2005}. 
In this setting, the large spin of our model represents an effective description of the collective nuclear spin system~\cite{LopezMonis_NJP_2011} and the electronic part provides a minimal model for investigating the effects on transport of coupled spin-spin dynamics.


A further potential realization of the large spin in our model is a magnetic impurity in a quantum dot.  
Several recent works have considered the influence of such an impurity on the transport properties through the dot \cite{Ramon07,Kiesslich_APL_2009,Elste10,Sothmann10}.  
In this context, our model can be viewed as the large-spin counterpart of the previously studied models and in particular the spin-1/2 impurity model of Refs.~[\onlinecite{Kiesslich_APL_2009}] and~[\onlinecite{Sothmann10}].
This possibility is also closely related to transport through single molecular magnets\cite{Heersche06,Bogani08,Friedman10,Bode_arXiv_2011}, for which our large spin would map to a magnetic atom and the isolated levels of our quantum dot to molecular orbitals.

We mention that our study of classical chaos in a quantum dot with coupling to an external pseudo-spin 
is also complementary to previous studies of intrinsic {\em quantum} chaos of, e.g.,  ballistic quantum dots. Those latter systems are often analyzed with statistical tools such as random matrix theory~\cite{Beenakker_RMP_1997,Alhassid_RMP_2000}.
 
The outline of this paper is as follows. In Section II, we introduce the model Hamiltonian and the equations of motion. Section III presents results and a classification of various non-linear regimes in the form of a map in parameter space, and we conclude with a brief discussion of the experimental relevance of our finding in Section IV.

\section{Model}

\subsection{Hamiltonian}

\begin{figure}[t]
  \centering
	\subfloat[]{\label{fig1a}\includegraphics[width=0.3\textwidth]{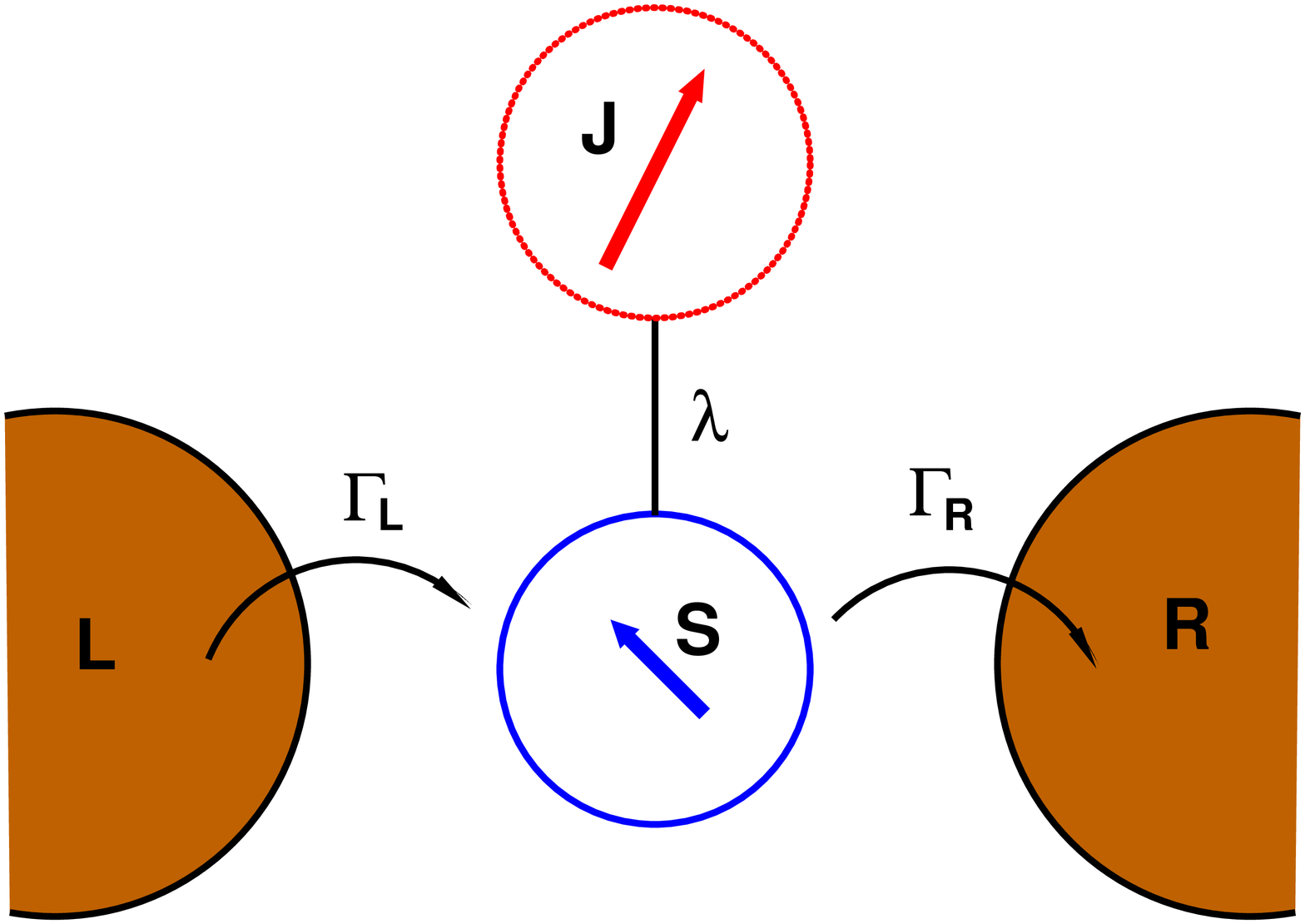}} \hspace{0.1cm}
	\subfloat[]{\label{fig1b}\includegraphics[width=0.3\textwidth]{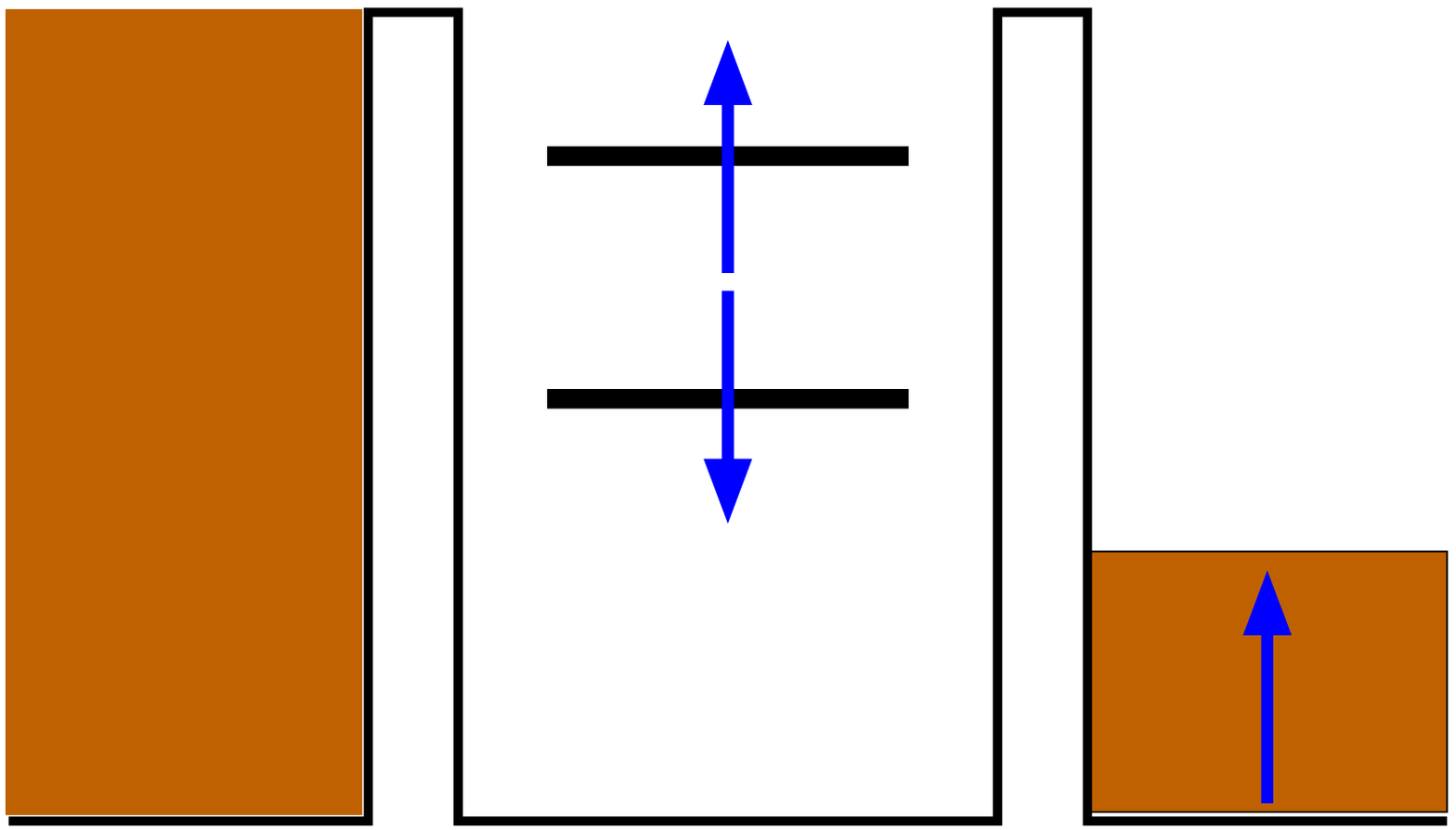}}
\caption{(Colour online.) Scheme and setup of the investigated system. a) An electron spin $\hat{\mathbf{S}}$ (blue arrow) in a QD is coupled via the exchange interaction $\lambda$ with a large spin $\hat{\mathbf{J}}$ (red arrow). The QD is attached to ferromagnetic electron reservoirs (brown regions), allowing electrons to tunnel through the QD. The large spin is isolated. b) The spin-dependent rates of the contact barriers are chosen so that a spin down electron is always trapped in the QD, while spin up electrons can tunnel through it (see details in the text). The large spin interacts with the spin of the electron trapped in the QD, allowing its spin to flip and, hence, escape form the QD into the right lead.}
  \label{qd}
\end{figure}

We investigate a quantum dot (QD) with a single orbital level, coupled to an emitter (left electron lead), a collector (right electron lead)  and to a {\it large} spin $\hat{\mathbf{J}}$ (Fig.~\ref{fig1a}). An external magnetic field $B_z$ is applied in $z$-direction which splits the QD spin levels (Fig.~\ref{fig1b}). 
The Hamiltonian for this model is
\begin{eqnarray} \label{ham}
\hat{H} & = & \hat{H}_{FA} + \hat{H}_J + \hat{V}.
\end{eqnarray}
Here, $\hat{H}_{FA}$ is the Fano-Anderson model for the QD coupled to the leads, which is exactly solvable; $\hat{H}_J$ is the Hamiltonian for the free motion of the large spin due the external magnetic field; and $\hat{V}$ is the coupling between a dot electron and the large spin. These individual Hamiltonians read:
\begin{subequations}
\begin{eqnarray}
\hat{H}_{FA} & = & \sum_{\sigma} \epsilon_d \hat{d}^{\dagger}_{\sigma} \hat{d}_{\sigma} + B_z \hat{S}_z \nonumber \\ & + & \sum_{lk \sigma} \, \epsilon_{lk \sigma} \hat{c}^{\dagger}_{lk \sigma} \hat{c}_{lk \sigma} + \sum_{lk \sigma} \, \left( \gamma_{lk} \hat{c}^{\dagger}_{lk \sigma} \hat{d}_{\sigma} + \text{h.c.}  \right) \, , \label{HFA} \\
\hat{H}_J & = & B_z \hat{J}_z \, , \label{HJ} \\
\hat{V} & = & \sum_{i = x, y, z} \lambda_i \hat{S}_i \hat{J}_i \, , \label{V}
\end{eqnarray}
\end{subequations}
where $\epsilon_d$ is the energy of the QD level, $\hat{d}_{\sigma}^{\dagger}$/$\hat{d}_{\sigma}$ creates/annihilates a spin-$\sigma$ electron in the dot, $\hat{S}_i$ is the $i$-th component of the electron spin operator in second quantization, $\hat{J}_i$ is the $i$-th component of large spin operator, and $\lambda_i$ is the coupling between the $i$-th components of the electron and the large spin, $\hat{c}_{lk\sigma}^{\dagger}$/$\hat{c}_{lk\sigma}$ creates/annihilates an electron with momentum $k$ and spin $\sigma$ in lead $l \in \{L,R\}$, and $\gamma_{lk}$ is the coupling between the QD and the $l$-th lead. Coulomb interaction in the QD is neglected, and thus double occupation is allowed. The flip-flop processes due to the spin-spin interaction are the origin of the non-trivial dynamics that will be shown in the next section. Much of the interesting dynamics found occurs at low magnetic fields, in particular, in a regime where the coupling between the electron and the large spin dominates Zeeman splittings ($B_z \ll \lambda$). Thus, in this regime we believe that different $g$-factors will not be qualitatively important, meaning the energy mismatch between the Zeeman splittings will not lead to suppression of the flip-flop processes. Therefore, for simplicity, we assume identical $g$-factors for the electron spin and the large spin, and absorb the Bohr magneton and the $g$-factors into the definition of $B_z$.


The classical counterpart of the closed system ($\gamma_{lk} = 0$) is, for zero external magnetic field ($B_z = 0$), a completely integrable system for arbitrary $\lambda_i$~\cite{Magyari_ZPhysB_1987}, while the isotropic model ($\lambda_x = \lambda_y = \lambda_z$) is also completely integrable for finite external magnetic fields. However, in presence of a finite magnetic field, an {\it anisotropic} coupling between the electron spin and the large spin, makes the model non-integrable and can lead to a chaotic spin dynamics~\cite{Robb_PRE_1998}. Therefore, in this work we take the coupling between the electron spin and the large spin to be anisotropic, and for simplicity we will focus on the choice
\begin{eqnarray}
\lambda_x = \lambda_z = \lambda, \quad  \lambda_y = 0.
\end{eqnarray}
Finally, the spin-dependent rates of the contact barriers are chosen so that only spin-up electrons can tunnel out of the QD
(Fig.~\ref{fig1b}), and $B_z \gg k_BT$ where $T$ is the temperature of the leads and $k_B$ is Boltzmann's constant. 
In this regime, current can flow only through the spin-up level of the QD. When an electron enters the spin-down level, it remains trapped until a spin-flip process (due to the interaction with the large spin) produces a transition from the spin-down to the spin-up level, allowing the trapped electron to escape the QD. 
Notice that because we have taken identical  $g$-factors for both the electron spin and the large spin, the spin-flip transition from the QD spin-down to the spin-up level conserves energy and the energy that the electron absorbs in the spin-flip is emitted by the large spin.


\subsection{Equations of Motion}

The equation of motion (EOM) for the expectation value of an operator $\hat{A}$ is
\begin{eqnarray} \label{heom}
\frac{d}{dt} \, \langle \hat{A} \rangle & = & \frac{1}{i \hbar} \langle [\hat{A},\hat{H}] \rangle + \left\langle \frac{\partial \hat{A}}{\partial t} \right\rangle.
\end{eqnarray}
Using this formula, we derive the EOM of each operator in Eq.~\eqref{HFA}, Eq.~\eqref{HJ} and Eq.~\eqref{V}. 

We first observe that the length of the large spin   $j = |\hat{\mathbf{J}}|$ is a conserved quantity since
$[\hat{\mathbf{J}}^2,\hat{H}] =0$. Next, due to the interaction $\hat{V}$, the EOMs do not close and lead to an infinite hierarchy of equations that needs to be truncated.  In order to do so, we use a factorization approximation by invoking a   
{\it mean-field} approximation for  $\hat{V}\to \hat{V}_{MF}$,
\begin{eqnarray} \label{VMF}
\hat{V}_{MF} & = & \sum_{i = x, y, x} \lambda_i \left( \hat{S}_i \langle \hat{J}_i \rangle + \hat{J}_i \langle \hat{S}_i \rangle - \langle \hat{S}_i \rangle \langle \hat{J}_i \rangle \right)
\end{eqnarray}
which neglects the  term $\delta \hat{S}_i \delta \hat{J}_i$ with 
$\delta \hat{S}_i = \hat{S}_i - \langle \hat{S}_i \rangle$ and $\delta \hat{J}_i = \hat{J}_i - \langle \hat{J}_i \rangle$, i.e. the quantum fluctuations of the electron spin and the external spin. We expect this to be a good approximation when $j\gg 1$ and the external spin $\hat{\mathbf{J}}$ can essentially be treated as a classical object due to its interaction with other environmental degrees of freedom. Furthermore, as in the semiclassical approximation we neglect quantum fluctuations of the large spin, we have no spin decay, meaning the large spin is a constant of motion.


We furthermore neglect terms proportional to $\gamma_{lk} \lambda_i$, namely, second order transitions due to the coupling of the large spin with the contacts. This is a good approximation in the infinite bias regime. For the electron leads, we perform the usual Born-Markov and flat band approximations and consider them to be in thermal equilibrium. Moreover, we consider the infinite bias regime, namely $\mu_{L} \to \infty$ and $\mu_R \to -\infty$, respectively (see Appendix A and B for details). The resulting EOMs read
\begin{eqnarray} \label{eoms}
\frac{d}{dt} \, \langle \hat{n}_{\sigma} \rangle & = & \lambda \langle \hat{J}_x \rangle \langle \hat{S}_y \rangle \left( \delta_{\sigma \uparrow} - \delta_{\sigma \downarrow} \right) - \Gamma \langle \hat{n}_{\sigma} \rangle + \Gamma_{L\sigma} \nonumber \\ 
\frac{d}{dt} \, \langle \hat{S}_x \rangle & = & -\left( \lambda \langle \hat{J}_z \rangle + B_z \right) \langle \hat{S}_y \rangle - \Gamma \langle \hat{S}_x \rangle \nonumber \\ 
\frac{d}{dt} \, \langle \hat{S}_y \rangle & = & -\lambda \langle \hat{J}_x \rangle \langle \hat{S}_z\rangle + \left( \lambda \langle \hat{J}_z \rangle +  B_z \right) \langle \hat{S}_x \rangle - \Gamma \langle \hat{S}_y \rangle \nonumber \\ 
\frac{d}{dt} \, \langle \hat{S}_z \rangle & = & \lambda \langle \hat{J}_x \rangle \langle \hat{S}_y \rangle - \Gamma \langle \hat{S}_z \rangle + \frac{1}{2}\left( \Gamma_{L\uparrow} - \Gamma_{L\downarrow} \right) \nonumber \\
\frac{d}{dt} \, \langle \hat{J}_x \rangle & = & -\left( \lambda \langle \hat{S}_z \rangle + B_z \right) \langle \hat{J}_y \rangle \nonumber \\ 
\frac{d}{dt} \, \langle \hat{J}_y \rangle & = & -\lambda \langle \hat{S}_x \rangle \langle \hat{J}_z\rangle + \left( \lambda \langle \hat{S}_z \rangle +  B_z \right) \langle \hat{J}_x \rangle \nonumber \\ 
\frac{d}{dt} \, \langle \hat{J}_z \rangle & = & \lambda \langle \hat{S}_x \rangle \langle \hat{J}_y \rangle,
\end{eqnarray}
where $\hat{n}_{\sigma} = \hat{d}_{\sigma}^{\dagger} \hat{d}_{\sigma}$, $\Gamma_{\sigma} = \Gamma_{L\sigma} + \Gamma_{R\sigma}$ with $\sigma = \uparrow, \downarrow$, and $\Gamma_{L\sigma}$ and $\Gamma_{R\sigma}$ are the tunnelling rates through the left and right contact barriers, respectively. We have taken $\Gamma_{\uparrow} = \Gamma_{\downarrow} = \Gamma$ for simplicity. In order to have current only through the spin-up level we take $\Gamma_{R\downarrow} = 0$. Therefore, spin-up electrons are allowed to tunnel through the QD, whereas spin-down electrons become trapped in it.

The EOM for the total number of electrons in the QD ($\hat{N} = \hat{n}_{\uparrow} + \hat{n}_{\downarrow}$) is independent of both the electron and the large spin components and is exactly solvable (see Appendix A). Thus, as $2\hat{S}_z = \hat{n}_{\uparrow} - \hat{n}_{\downarrow}$, the level occupations can be obtained through the following expression:
\begin{eqnarray} \label{ocu}
\langle \hat{n}_{\sigma} (t) \rangle & = & \frac{1}{2} \left( \langle \hat{N} (0) \rangle \, e^{-\Gamma t} + \frac{\Gamma_{L\uparrow} + \Gamma_{L\downarrow}}{\Gamma} \left( 1 - e^{-\Gamma t} \right) \right) \nonumber \\ & + & (\delta_{\sigma \uparrow} - \delta_{\sigma \downarrow}) \langle \hat{S}_z  (t) \rangle
\end{eqnarray}
which relates the level occupation with the $z$-component of the electron spin. 
If the coupling between the electron and the large spins is isotropic ($\lambda_x = \lambda_y = \lambda_z$), it is straightforward to see that in the stationary limit the spins decouple, and the well known Fano-Anderson solution is obtained (see Appendix C). In contrast, we show below that the situation is drastically different for the anisotropic case where the stationary solutions for the EOMs depend on the coupling between the spins.

The average electron current $\langle \hat{I} \rangle$ through the QD is solely due to a decay at rate $\Gamma_{R\uparrow}$ from the spin-up QD level into the right lead,
\begin{eqnarray}
\langle \hat{I}(t) \rangle = e \Gamma_{R\uparrow} \langle \hat{n}_{\uparrow}(t) \rangle,
\end{eqnarray}
where $e$ denotes the electron charge. In the long-time limit of the current can be written as (see Eq.~\eqref{ocu})
\begin{eqnarray} \label{crnt}
\frac{\langle \hat{I}(t) \rangle}{e \Gamma_{R\uparrow}} & = & \frac{1}{2} \frac{\Gamma_{L\uparrow} + \Gamma_{L\downarrow}}{\Gamma} + \langle \hat{S}_z(t) \rangle.
\end{eqnarray}
Henceforth, for convenience we take $\Gamma_{L\uparrow} = \Gamma_{R\uparrow} = \Gamma/2$. Other options give similar behaviour except for the transient solutions. 

\section{Regions in parameter space}

\begin{figure}
\includegraphics[width=0.5\columnwidth]{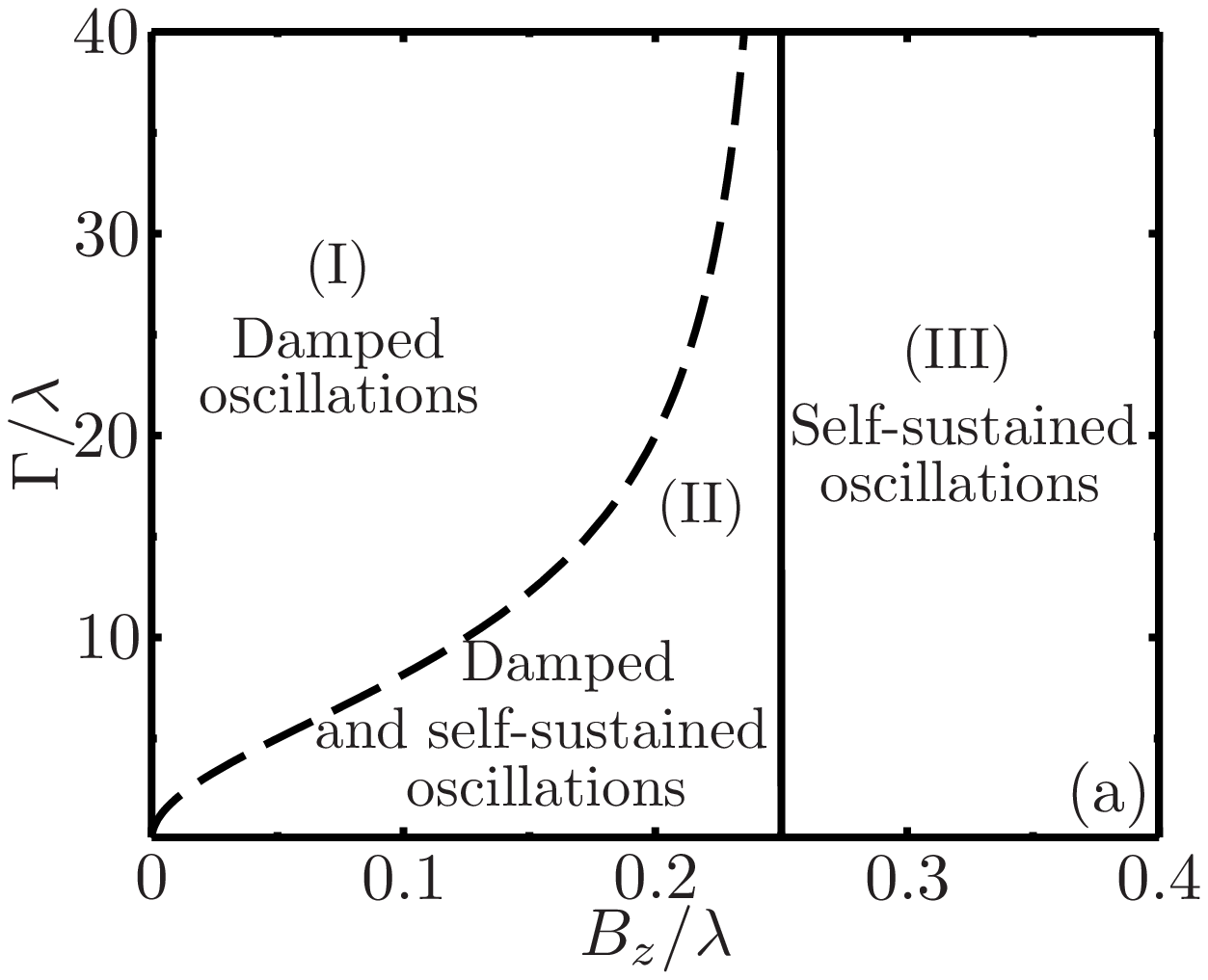}
\includegraphics[width=0.42\columnwidth]{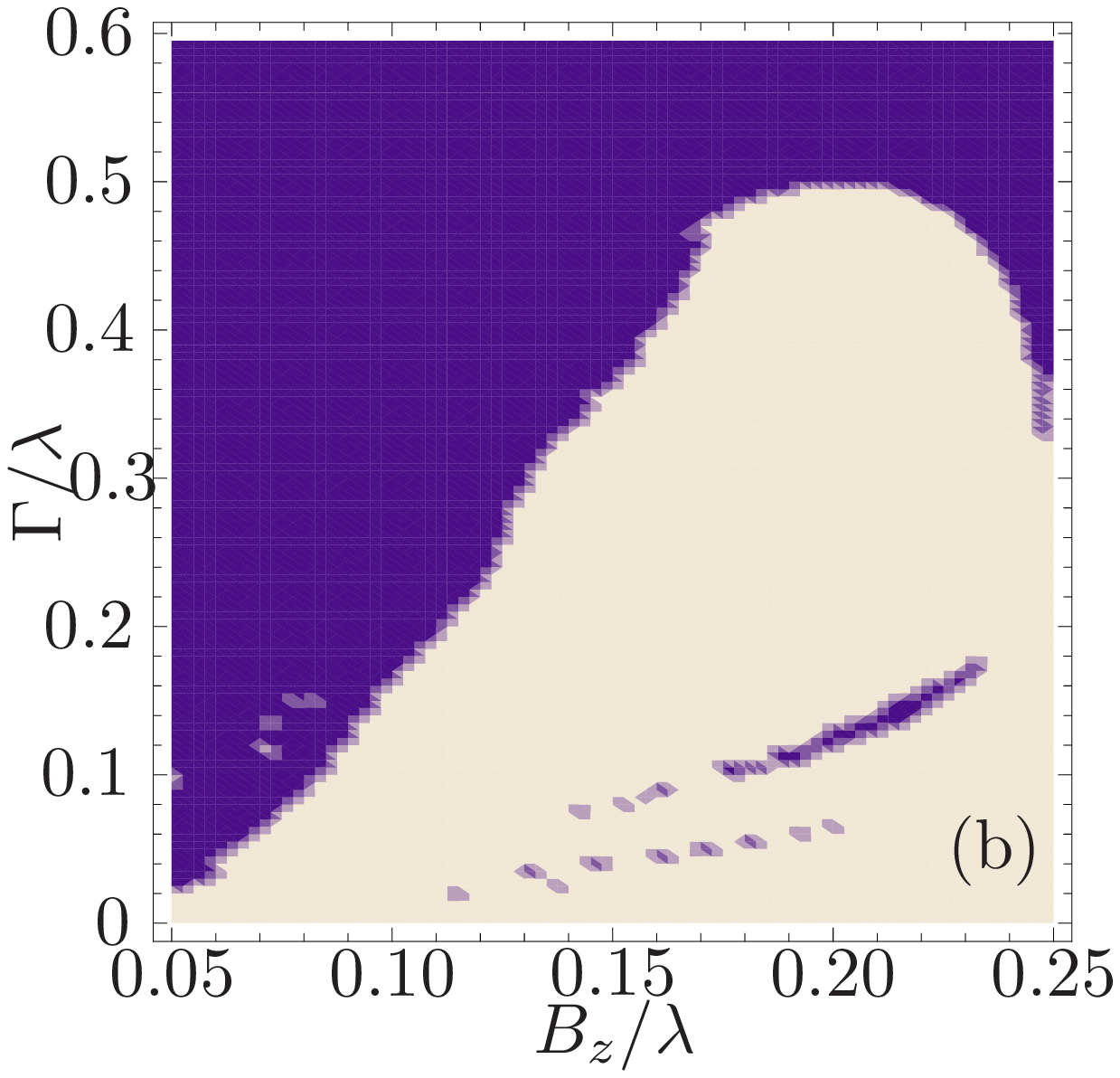}
\caption{a) Parameter space with three regions describing the behaviour of solutions of EOMs (Eq. (6)). Boundaries between the regions are obtained analytically from Eq. (14). Region I: damped oscillations; region II: both damped and self-sustained oscillations; and region III: self-sustained oscillations only.
b) Numerically-obtained small-$\Gamma$ region in the mixed region II. In the dark-coloured region, damped oscillations are obtained.  In the light-colored region: self-sustained oscillations.
Initial conditions: $\langle \hat{S}_x \rangle_{t = 0} = 1/2$ and $\langle \hat{S}_y \rangle_{t = 0} = \langle \hat{S}_z\rangle_{t = 0} = 0$, $\langle \hat{J}_x \rangle_{t = 0} = \langle \hat{J}_y \rangle_{t = 0} = j/2$ and $\langle \hat{J}_x \rangle_{t = 0} = j/\sqrt{2}$.}
\label{fig2}
\end{figure}

The stationary solutions of the EOMs, Eq.~\eqref{eoms}, can be obtained analytically and we find eight fixed points. Two of these fixed points, however,  always have a finite imaginary component, and as they have no physical meaning, we leave them out of the subsequent analysis.   The remaining fixed points serve to divide the parameter space of the model into distinct regions, as shown in Fig.~\ref{fig2}.

Introducing the notation 
\begin{eqnarray}
\mathcal{P} & = & \left( \langle \hat{S}_x \rangle, \langle \hat{S}_y \rangle, \langle \hat{S}_z \rangle,\langle \hat{J}_x \rangle, \langle \hat{J}_y \rangle, \langle \hat{J}_z \rangle \right),
\end{eqnarray}
the six relevant fixed points are
\begin{subequations}
\begin{eqnarray}
\mathcal{P}_{\pm} & = & \left( 0,0, -\frac{1}{4},0,0,\pm j \right), \label{PMm} \\ \mathcal{P}_{\mathrm{II},1_{\pm}} & = & \left( 0, \mathcal{B}_2, -\frac{B_z}{\lambda}, \frac{\Gamma}{B_z} \, \mathcal{B}_2, \pm \mathcal{B}_1, -\frac{B_z}{\lambda} \right) \label{PII1}, \\ \mathcal{P}_{\mathrm{II},2_{\pm}} & = & \left( 0, -\mathcal{B}_2, -\frac{B_z}{\lambda}, -\frac{\Gamma}{B_z} \, \mathcal{B}_2, \pm \mathcal{B}_1, -\frac{B_z}{\lambda} \right) \label{PII2},
\end{eqnarray}
\end{subequations}
where
\begin{subequations}
\begin{eqnarray}
\mathcal{B}_1 & = & -\sqrt{ j^2 - \left( \frac{\lambda}{4 B_z} - 1 \right) \left( \frac{\Gamma}{\lambda} \right)^2 - \left( \frac{B_z}{\lambda} \right)^2}, \label{B1} \\ \mathcal{B}_2 & = & -\sqrt{\frac{B_z}{\lambda} \left( \frac{1}{4} - \frac{B_z}{\lambda} \right)}. \label{B2}
\end{eqnarray}
\end{subequations}
For certain values of $B_z$, $\Gamma$ and $\lambda$, the quantities $\mathcal{B}_1$ and $\mathcal{B}_2$ (\eqref{B1} and Eq.~\eqref{B2}) can have finite imaginary components and therefore, points $\mathcal{P}_{\mathrm{II},1_{\pm}}$ and $\mathcal{P}_{\mathrm{II},2_{\pm}}$ only have physical meaning in the region of parameter space where $\mathcal{B}_1$ and $\mathcal{B}_2$ are real. 
Fig.~\ref{fig2}a shows a projection of the 3-dimensional parameter space on the $\Gamma$ versus $B_z$ plane for a fixed $\lambda$. This diagram is divided in three regions. 
In region I, $\mathcal{B}_1$ is a pure imaginary number, and hence, $\mathcal{P}_{\mathrm{II},1_{\pm}}$ and $\mathcal{P}_{\mathrm{II},2_{\pm}}$ are nonphysical, and $\mathcal{P}_{\pm}$ the only physical fixed points.
In region II, $\mathcal{B}_1$ and $\mathcal{B}_2$ are both real, and all six fixed points are physical. 
In region III, $\mathcal{B}_2$ is purely imaginary,  and again $\mathcal{P}_{\pm}$ are the only physical fixed points. 
Points $\mathcal{P}_{\pm}$ are thus physical solutions for the EOMs Eq.~\eqref{eoms} in all three regions, whereas the fixed points $\mathcal{P}_{\mathrm{II},\pm}$ are physical only in region II. The boundaries between the regions are obtained by solving the equations $\mathcal{B}_1 = 0$ and $\mathcal{B}_2 = 0$, namely
\begin{eqnarray} \label{bndry}
\mathcal{B}_1 = 0 & \Rightarrow & \Gamma = \sqrt{\frac{j^2 - (B_z/\lambda)^2}{1/4 - B_z/\lambda} \frac{B_z}{\lambda},}\nonumber \\
\mathcal{B}_2 = 0 & \Rightarrow & \frac{B_z}{\lambda} = \frac{1}{4},
\end{eqnarray}
and these two equations give the lines plotted in Fig.~\ref{fig2}a.



\subsection{Region I}

\begin{figure}
\includegraphics[width=0.9\columnwidth]{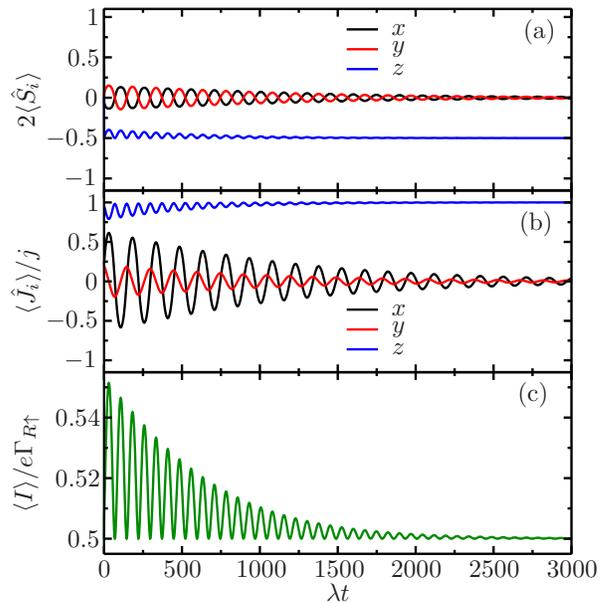}
  \caption{(Colour online.) Time evolution in region I of a) the electron spin components, b) the large spin components, and c) the current through the QD obtained by solving numerically the EOMs of Eq.~\eqref{eoms}. 
  In this region, the solutions exhibit a slow damped behaviour. In the stationary limit, the large spin is completely polarized in the direction parallel to the external magnetic field, and a QD electron trapped in the spin-down state. Current is due only to tunnelling through the spin up level and in the stationary limit tends to a constant value of $5/8$. The parameters here are $B_z/\lambda = 0.1$ and $\Gamma/\lambda = 9$, with initial conditions $\langle \hat{S}_x \rangle_{t = 0} = 1/2$, $\langle \hat{S}_y \rangle_{t = 0} = \langle \hat{S}_z \rangle_{t = 0} = 0$, $\langle \hat{J}_x \rangle_{t = 0} = \langle \hat{J}_y \rangle_{t = 0} = (5/\sqrt{2})(\sqrt{5} - 1)/2$ and $\langle \hat{J}_z \rangle_{t = 0} =  (5/\sqrt{2})\sqrt{5 + \sqrt{5}}$.}
    \label{fig3}
\end{figure}

In order to obtain the time evolution of the electron and large spin components and the electronic current through the QD, the EOMs Eq.~\eqref{eoms} are solved numerically.
Fig.~\ref{fig3} shows the time evolution of the electron spin and the large spin components, and the current through the QD in region I of the parameter space.
All exhibit completely damped oscillations. In the previous discussion, we have seen that in region I, $\mathcal{P}_{\pm}$ Eq.~\eqref{PMm} are the only physical fixed points. Depending on the choice of parameters and initial conditions, the system will evolve to $\mathcal{P}_+$ or $\mathcal{P}_-$.  For the parameters and initial conditions chosen in Fig.~\ref{fig3}, the system evolves towards the fixed point $\mathcal{P}_+$. In this case, the large spin becomes completely polarized in the direction parallel to the external magnetic field (Fig.~\ref{fig3}b), and a spin-down electron remains trapped in the QD (Fig.~\ref{fig3}a) and the interaction between the electron spin and the large spin is no longer effective. Spin-up electrons, however, can still tunnel through the QD (Fig.~\ref{fig3}c), and in the stationary limit the current becomes (see Eq.~\eqref{crnt})
\begin{eqnarray} \label{crntI}
\frac{\langle \hat{I}(t) \rangle}{e \Gamma_{R\uparrow}} & = & \frac{1}{2}.
\end{eqnarray}
In region I, then, the coupling of the two spin systems with the external leads results in complete damping of the transient oscillations of the electron and the large spin components and the current. 
A finite, fully spin-polarized electron current flows through the QD that in the stationary limit is not influenced by the interaction with the large spin.

\subsection{Region II}

\begin{figure*}
\includegraphics[width=0.9\textwidth]{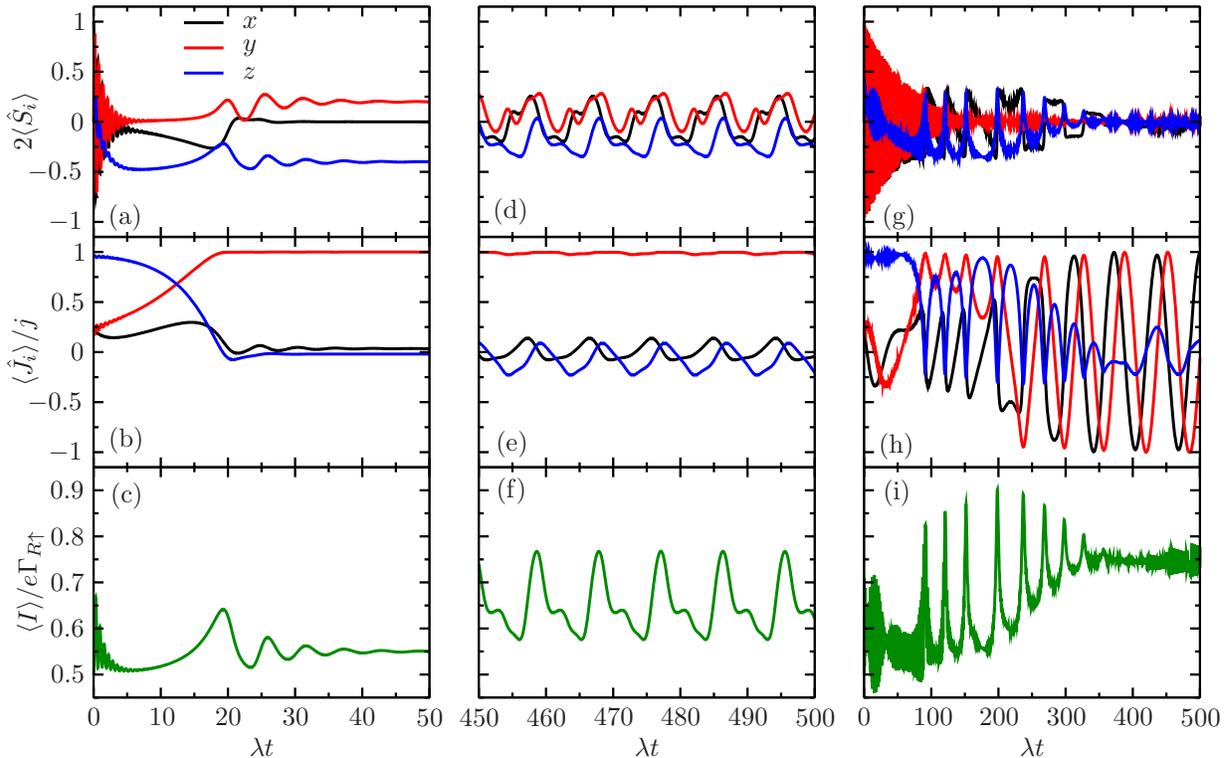}
  \caption{Results for three different parameter sets are shown, each of which gives rise to very different system behaviour.
    Figures a), b), c) show fast damping behaviour ($B_z/\lambda = 0.2$, $\Gamma/\lambda = 0.7$, dark region in Fig.~\ref{fig2}b). In the long-time limit the large spin is almost completely polarized in the direction perpendicular to the external field, but unlike in region I, the spin down electron can escape from the QD into the right lead due to the interaction with the large spin. 
    Figures d), e), f) show periodic self-sustained oscillations ($B_z/\lambda = 0.1$, $\Gamma/\lambda = 0.16$, light region in Fig.~\ref{fig2}b), which is a signature of limit-cycles in phase space
    (see Fig.~\ref{fig5}). 
    Figures g), h), i) show chaotic self-sustained oscillations ($B_z/\lambda = 0.1$, $\Gamma/\lambda = 0.015$, light region in Fig.~\ref{fig2}b). 
    In the oscillatory cases, the oscillations are captured in the current through the QD, and in particular, the chaotic behaviour is observed in the current (Fig.~\ref{fig4}i). The initial conditions are $\langle \hat{S}_x \rangle_{t = 0} = 1/2$, $\langle \hat{S}_y \rangle_{t = 0} = \langle \hat{S}_z \rangle_{t = 0} = 0$, $\langle \hat{J}_x \rangle_{t = 0} = \langle \hat{J}_y \rangle_{t = 0} = (5/\sqrt{2})(\sqrt{5} - 1)/2$ and $\langle \hat{J}_z \rangle_{t = 0} =  (5/\sqrt{2})\sqrt{5 + \sqrt{5}}$.}  
  \label{fig4}
\end{figure*}

In region II,
the EOMs Eq.~\eqref{eoms} exhibit both damped and self-sustained oscillatory solutions, depending on the choice of parameters and initial conditions. Fig.~\ref{fig2}b shows the part of region II where the self-sustained oscillations are found. This behaviour can be seen for all intensities of the external magnetic field in region II, but only for small values of coupling $\Gamma$ with the leads. Comparing Fig.~\ref{fig2}a and Fig.~\ref{fig2}b we see that most values for $B_z$ and $\Gamma$  in region II lead to damped oscillations. Furthermore, although we have given analytical expressions for the boundaries between the different regions Eq.~\eqref{bndry}, we have not found an expression for the boundary between the regions inside region II where self-sustained and damped oscillations are found. 
Fig.~\ref{fig2}b has been obtained by solving the EOMs Eq.~\eqref{eoms} in region II. As can be seen, the boundary between both regions is fuzzy in contrast with the ones obtained between regions I, II and III Eq.~\eqref{bndry}. Moreover, Fig.~\ref{fig2}b shows small ``islands'' in the oscillatory region, where damped solutions are obtained. 
\subsubsection{Damped Oscillations}

Fig.~\ref{fig4}a, Fig.~\ref{fig4}b and Fig.~\ref{fig4}c, show the time evolution of the electron and the large spin components, and the current in region II with parameters $B_z$ and $\Gamma$ such that they all exhibit damped oscillations. Previously we have seen that in region II all the six fixed points are physical. For the parameters and initial conditions chosen in Fig.~\ref{fig4}a, Fig.~\ref{fig4}b and Fig.~\ref{fig4}c, the system evolves towards the fixed point $\mathcal{P}_{{\rm II},1+}$. The large spin becomes almost completely polarized in the $y$-direction (Fig.~\ref{fig4}b), perpendicular to the external magnetic field. 
and the current becomes (see Eq.~\eqref{crnt})
\begin{eqnarray}
\frac{\langle \hat{I} \rangle}{e \Gamma_{R\uparrow}} & = & \frac{3}{4} - \frac{B_z}{\lambda}
.
\end{eqnarray}
Thus, the stationary current increases if either the external magnetic field decreases or the coupling between the spins increases. Since in region II $B_z/\lambda < 1/4$, the coupling between the electron and the large spins enhances the current through the QD, compared with the current obtained in region I (Eq.~\eqref{crntI}). Nevertheless, the result of coupling the two spins to the leads stills yields complete damping of both spin oscillations, as in region I.

\subsubsection{Self-Sustained Oscillations and Chaos}

\begin{figure}
\includegraphics[width=0.9\columnwidth]{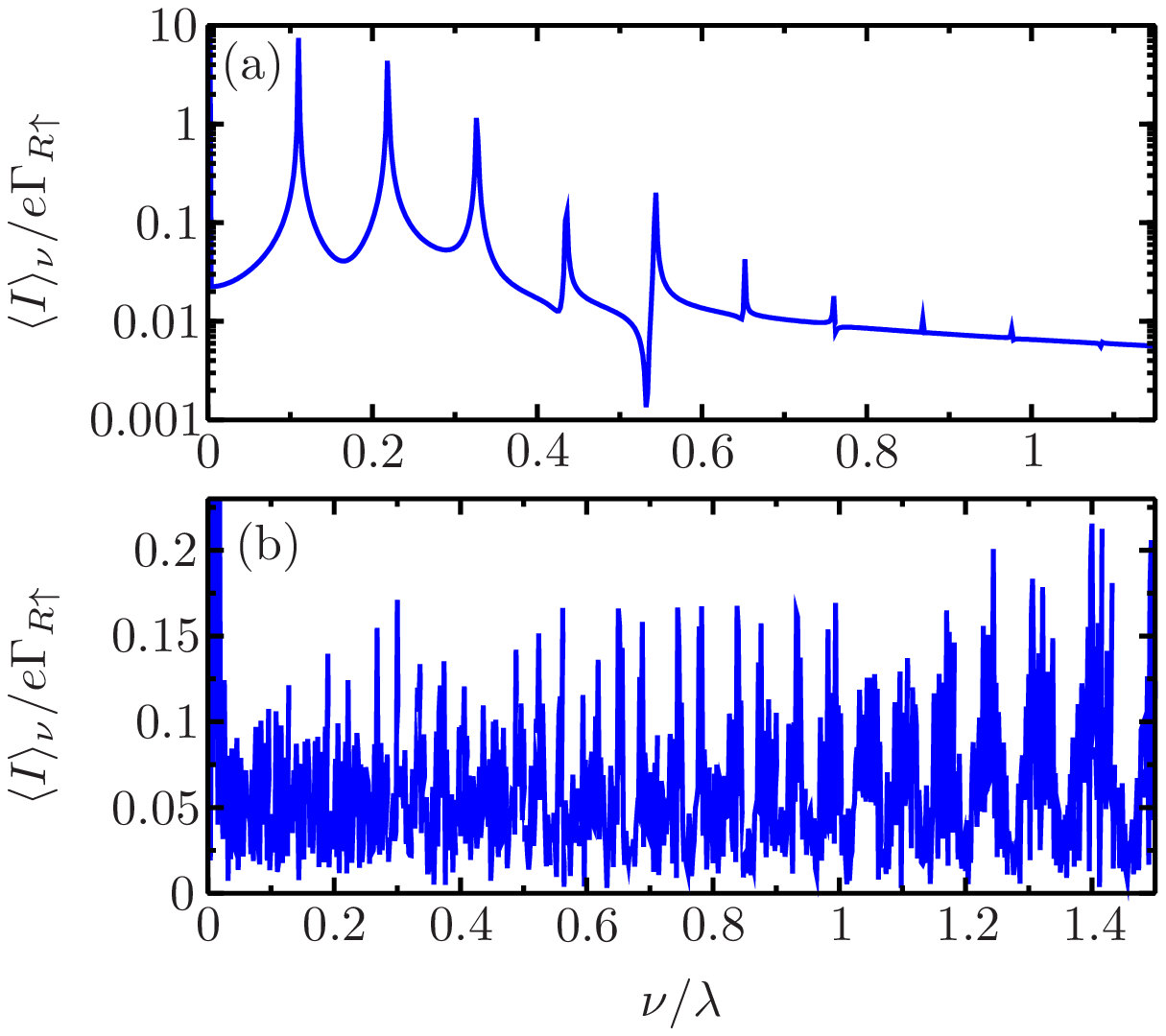}
  \caption{Fourier spectra of the non-damped current time evolutions shown in Fig.~\ref{fig4} in the long-time limit. Figs. a) and b) show the Fourier transform of Fig~\ref{fig4}f and~\ref{fig4}i, respectively, where $\nu$ is the frequency. Figure a) shows peaks at well defined frequencies, meaning that behaviour of the current is periodic. However, figure b) shows a uniform frequency distribution, which is a signature of chaotic dynamics.\vspace{0.5cm}}
\label{fig5}
\end{figure}

\begin{figure}
\includegraphics[width=0.95\columnwidth]{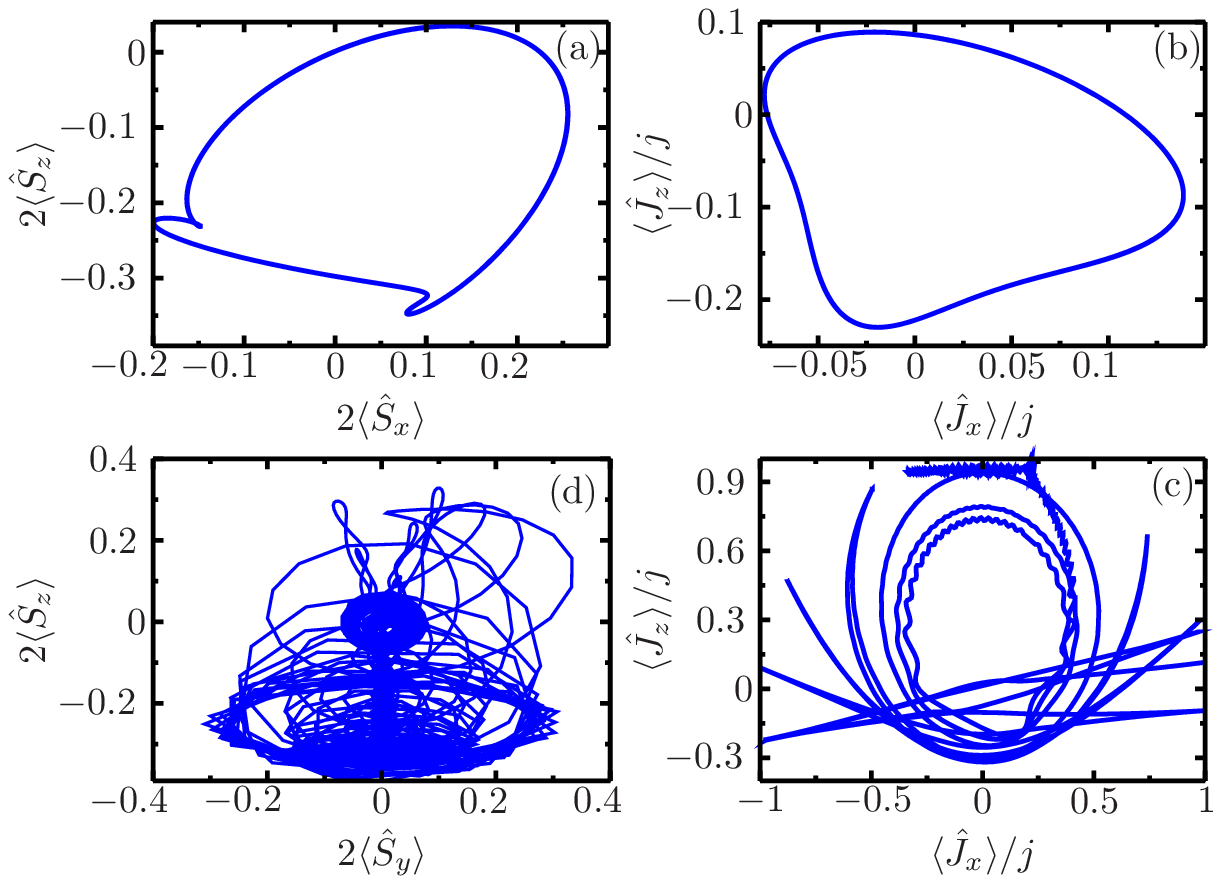}
  \caption{Electron spin (left figures) and large spin (right) trajectories projected on a two dimensional plane for the non-damped solutions in region II (light region in Fig.~\ref{fig2}a). 
  Figures a) and b) show the formation of a limit-cycle as seen in the time evolution plots,  Fig.~\ref{fig4}d, Fig.~\ref{fig4}e, and Fig.~\ref{fig4}f ($B_z/\lambda = 0.1$, $\Gamma/\lambda = 0.16$).
  Figures c) and d) correspond to the time evolution plots of Fig.~\ref{fig4}g, Fig.~\ref{fig4}h, and Fig.~\ref{fig4}i which suggest that the trajectories are chaotic ($B_z/\lambda = 0.1$, $\Gamma/\lambda = 0.015$).}
  \label{fig6}
\end{figure}

We shall now focus on the small region in region II where self-sustained oscillatory solutions are found (Fig.~\ref{fig2}b). Fig.~\ref{fig4}d, Fig.~\ref{fig4}e and Fig.~\ref{fig4}f, show the time evolution of the electron and the large spin components, and the current through the QD. The chosen values of $B_z$ and $\Gamma$ lead to complicated, but periodic, undamped oscillations. Fig.~\ref{fig5}a shows the Fourier spectrum of the current time evolution of Fig.~\ref{fig4}f in the long-time limit. The spectrum exhibits peaks at well defined frequencies, which clearly confirms the periodic behaviour of the current. Furthermore, in non-linear systems, self-sustained oscillations are a signature of {\it limit-cycles} and in Fig.~\ref{fig5}a and Fig.~\ref{fig5}b we plot the electron and the large spin trajectories in phase space, projected on the $x$-$z$ plane, in the long-time limit. These figures show that the spin trajectories are precisely limit-cycles. For all the initial conditions chosen, the system always converges to them. Finally, Figs.~\ref{fig4}g, \ref{fig4}h and~\ref{fig4}i show that decreasing $\Gamma$ turns the periodic self-sustained oscillations chaotic. In this case, the Fourier spectrum of the current, shown in Fig.~\ref{fig5}b, is uniformly distributed through all frequencies, which is a clear signature of chaos. Fig.~\ref{fig5}c and Fig.~\ref{fig5}d show the electron and large spin trajectories in the long-time limit, where it can be seen that they perform complicated non-periodic paths. In this area of region II, the coupling between the interacting spins and the leads does not produce damping of the spins as in the previous cases. Moreover, the electron current through the QD captures the complicated dynamics due to the interaction between the electron and the large spin, as seen in Fig.~\ref{fig4}f and Fig.~\ref{fig4}i.



\subsection{Region III}

\begin{figure}
\includegraphics[width=0.9\columnwidth]{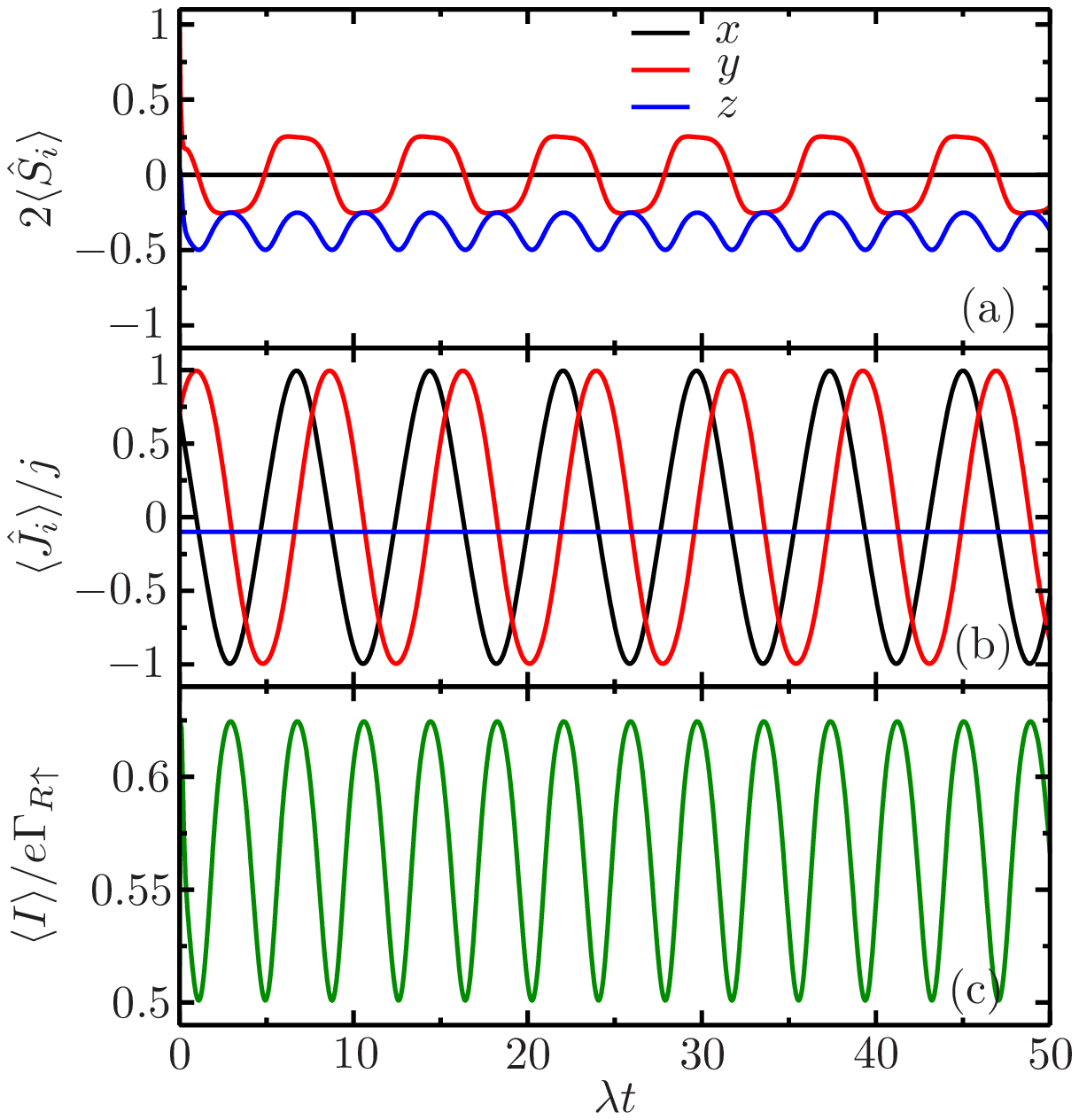}
  \caption{In this region the solutions exhibit periodic self-sustained oscillations, which 
  are reflected in the current.  The corresponding limit-cycles  are shown in Fig.~\ref{fig7}). The parameters chosen here are $B_z/\lambda = 1.0$, $\Gamma/\lambda = 10$. The initial conditions are $\langle \hat{S}_y \rangle_{t = 0} = 1/2$, $\langle \hat{S}_x \rangle_{t = 0} = \langle \hat{S}_z \rangle_{t = 0} = 0$, $\langle \hat{J}_x \rangle_{t = 0} = \langle \hat{J}_y \rangle_{t = 0} = 3\sqrt{11/2}$ and $\langle \hat{J}_z \rangle_{t = 0} = -1$.}  
  \label{fig7}
\end{figure}

\begin{figure}
\includegraphics[width=\columnwidth]{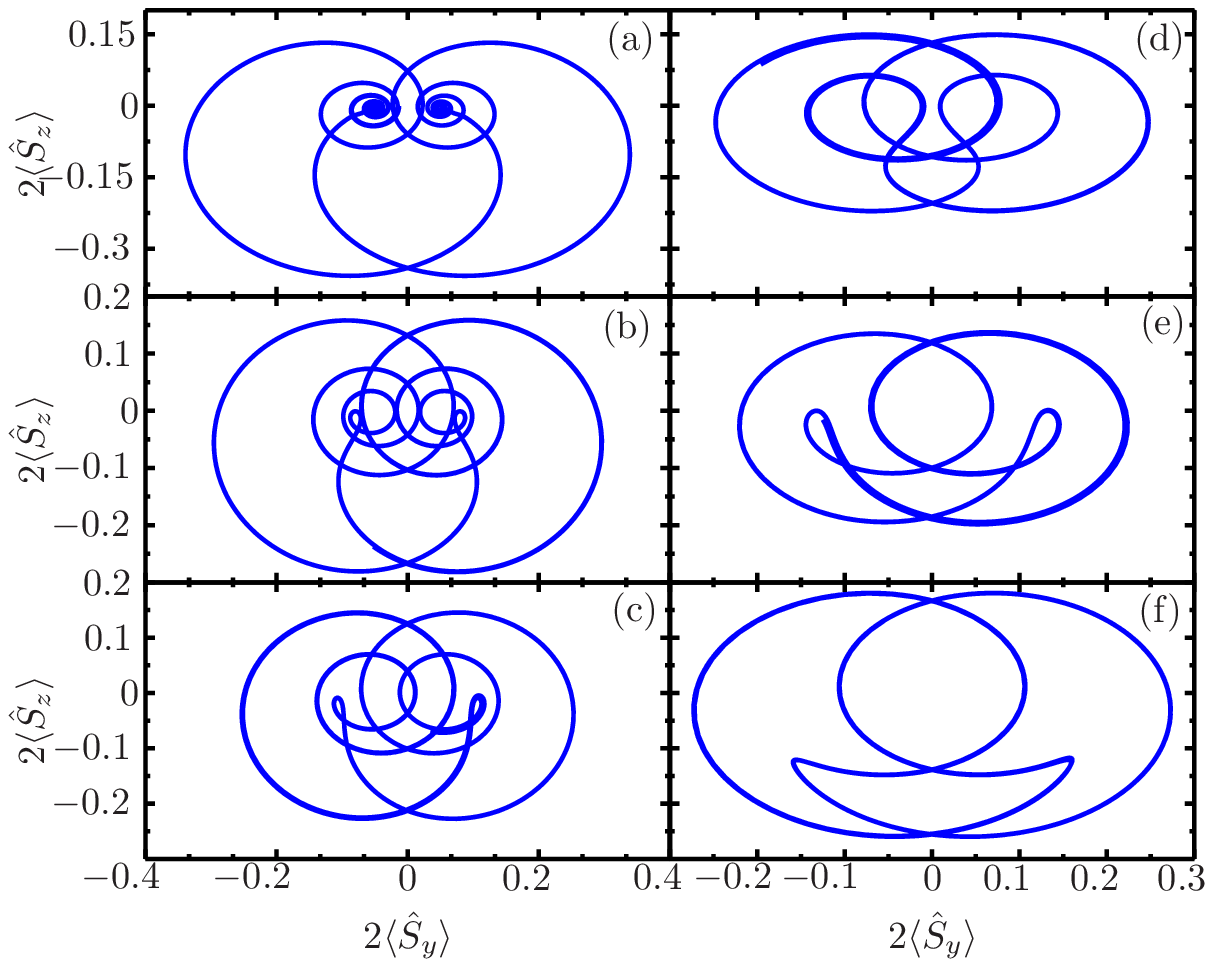}
  \caption{Electron spin trajectories projected in the $\langle \hat{S}_y \rangle$-$\langle \hat{S}_z \rangle$ plane in region III. The solutions of the EOMs given in Eq.~\eqref{eoms} are periodic self-sustained oscillations (Fig.~\ref{fig7}). Figures a)-f) show the different limit-cycles obtained when varying the external magnetic field. $\Gamma/\lambda = 1$.}
  \label{fig8}
\end{figure}

Fig.~\ref{fig8}a, Fig.~\ref{fig8}b and Fig.~\ref{fig8}c show the time evolution of the spin components and current for typical parameters in region III.
They all exhibit periodic self-sustained oscillations. 
Fig.~\ref{fig8} shows the different limit-cycles performed by the electron spin in phase space, projected in the $y$-$z$ plain, when the value of the external magnetic field is increased. 
The trajectories found for the large spin in the long-time limit suggest that this behaviour can be understood by means of an effective model in which the large spin simply acts on the QD electrons as an ac magnetic field in $x$-direction with amplitude
\begin{eqnarray}
B_{\rm ac}(t) & = & \frac{\lambda j}{\sqrt{2}} \left( \cos(B_zt) - \sin(B_zt) \right).
\end{eqnarray}
The EOMs for this effective model are (see Appendix D for details)
\begin{eqnarray} \label{effeoms}
\frac{d}{dt} \, \langle \hat{S}_x \rangle & = & -B_z \langle \hat{S}_y \rangle - \Gamma \langle \hat{S}_x \rangle \nonumber \\ \frac{d}{dt} \, \langle \hat{S}_y \rangle & = & B_z \langle \hat{S}_x \rangle - B_{\rm ac}(t) \langle \hat{S}_z \rangle - \Gamma \langle \hat{S}_y \rangle \nonumber \\ \frac{d}{dt} \, \langle \hat{S}_z \rangle & = & B_{\rm ac}(t) \langle \hat{S}_y \rangle - \Gamma \langle \hat{S}_z \rangle - \frac{\Gamma}{4}.
\end{eqnarray}
Thus, in this region the six autonomous non-linear equations Eq.~\eqref{eoms} can be approximated  by a set of three non-autonomous linear equations Eq.~\eqref{effeoms}. The agreement between the solutions obtained with this effective model and the full EOMs is very good.

In region III, the coupling between the two spin system leads to  self-sustained oscillations which are visible in the electron current through the QD, as shown in Fig.~\ref{fig8}c.

\section{Conclusions}


We have studied electron transport through a quantum dot coupled to ferromagnetic leads, in which the electron spin interacts with a large spin while an external magnetic field is applied. 
We have found that the motion of the electron spin, the large spin and the current through the QD strongly depend on the coupling between spins. When the electron spin and the large spin are isotropically coupled, the large spin becomes completely polarized and decouples from the electron spin. Conversely, when the electron spin and the large spin are anisotropicaly coupled, we have found that their motion and the current through the QD can either behave as in the isotropic case or show self-sustained oscillations which, furthermore, can be periodic or chaotic. Switching between different behaviours can be obtained by varying either the strength coupling with the leads or the intensity of the external magnetic field.


We foresee two possible experimental realisations of the large spin of our model.  The first is as an effective model on a hyperfine bath.  Here a semi-classical treatment may be justified by considering that the number of nuclei spins in semiconductor QDs interacting with an electron spin is very large (e.g., for GaAs QDs there are typically $10^5$-$10^6$ nuclei spins). 
Situations in which the hyperfine interaction is anisotropic have been discussed in Refs.~[\onlinecite{Coish09,Hodges08,Fischer09}].
The second realisation is that our large spin represents the spin of a magnetic impurity of a doped semiconductor or a magnetic atom in a single molecular magnet.  While in this case the spin may not be so large, mean-field analyses such as pursued here can still provide useful information, e.g. Ref.~[\onlinecite{Henelius_PRB_2008}].


From the theoretical point of view, it would be interesting to investigate how the features of this semiclassical treatment are reflected in a quantum master equation approach, in which the electron and the large spin are both treated as quantum objects.  This opens a path to investigate the quantum/classical divide in a nonequillibrium context.

\begin{acknowledgments}
The authors would like to thank S. Kohler, A. Metelmann, R. S\'anchez and G. Schaller for helpful discussions. We acknowledge financial support through Grant No. MAT2008-02626 (MICINN), from FPU grant (C. L\'opez-Mon\'is), from ITN under Grant No. 234970 (EU) and through Grant No. DE2009-0074 (DAAD-MICINN).
\end{acknowledgments}

\appendix

\begin{widetext}

\section{Derivation of the equations of motion}

In this appendix, we summarize the steps in the derivation of the EOMs (Eq.~\eqref{eoms}).  
We start with the Hamiltonian of Eq.~\eqref{ham} and, for later convenience, we shift the reservoir frequencies 
$ 
  \sum_{lk \sigma} \, \epsilon_{lk \sigma} \hat{c}^{\dagger}_{lk \sigma} \hat{c}_{lk \sigma} 
  \to 
  \sum_{lk \sigma} \, ( \epsilon_{lk \sigma} + \mu_{l}) \hat{c}^{\dagger}_{lk \sigma} \hat{c}_{lk \sigma} 
$ where $\mu_{l}$ is the chemical potential of lead $l$.  Under the mean-field approximation considered in this work (see Eq.~\eqref{VMF}), the closed set of EOMs obtained for the time evolution of operators in the Hamiltonian (Eq.~\eqref{ham}) are then computed to be:
\begin{eqnarray} \label{elec}
i \, \frac{d}{dt} \, \langle \hat{d}_{\sigma}^{\dagger} \hat{d}_{\sigma'} \rangle
& = & \frac{\lambda}{2} \left( \delta_{\sigma' \uparrow} \langle \hat{d}_{\sigma}^{\dagger} \hat{d}_{\downarrow} \rangle + \delta_{\sigma' \downarrow} \langle \hat{d}_{\sigma}^{\dagger} \hat{d}_{\uparrow} \rangle - \delta_{\sigma \uparrow} \langle \hat{d}_{\downarrow}^{\dagger} \hat{d}_{\sigma'} \rangle - \delta_{\sigma \downarrow} \langle \hat{d}_{\uparrow}^{\dagger} \hat{d}_{\sigma'} \rangle \right) \langle \hat{J}_x \rangle \nonumber \\ & + & \frac{1}{2} \left( \delta_{\sigma' \uparrow} \, \langle \hat{d}_{\sigma}^{\dagger} \hat{d}_{\uparrow} \rangle - \delta_{\sigma' \downarrow} \, \langle \hat{d}_{\sigma}^{\dagger} \hat{d}_{\downarrow} \rangle - \delta_{\sigma \uparrow} \, \langle \hat{d}_{\uparrow}^{\dagger} \hat{d}_{\sigma'} \rangle + \delta_{\sigma \downarrow} \, \langle \hat{d}_{\downarrow}^{\dagger} \hat{d}_{\sigma'} \rangle \right) \left( \lambda \langle \hat{J}_z \rangle + B_z \right) \nonumber \\ & - & \sum_{l, \, k} \, \left( \gamma_{lk} \langle \hat{c}_{lk \sigma}^{\dagger} \hat{d}_{\sigma'} \rangle - \gamma_{lk}^* \langle \hat{d}_{\sigma}^{\dagger} \hat{c}_{lk \sigma'}
\rangle \right)\\
i \, \frac{d}{dt} \, \langle \hat{c}_{lk \sigma}^{\dagger} \hat{d}_{\uparrow} \rangle & = & \frac{\lambda}{2} \langle \hat{J}_x \rangle \langle \hat{c}_{lk \sigma}^{\dagger} \hat{d}_{\downarrow} \rangle + \frac{1}{2} \left( \lambda \langle \hat{J}_z \rangle + B_z \right) \langle \hat{c}_{lk \sigma}^{\dagger} \hat{d}_{\uparrow} \rangle + \sum_{l', \, k'} \, \gamma_{l'k'}^* \langle \hat{c}_{lk \sigma}^{\dagger} \hat{c}_{l'k' \uparrow} \rangle - (\epsilon_{l k \sigma} + \mu_l) \langle \hat{c}_{l k \sigma}^{\dagger} \hat{d}_{\uparrow} \rangle - \gamma_{l k}^* \langle \hat{d}_{\sigma}^{\dagger} \hat{d}_{\uparrow} \rangle \nonumber \\
i \, \frac{d}{dt} \, \langle \hat{c}_{lk \sigma}^{\dagger} \hat{d}_{\downarrow} \rangle & = & \frac{\lambda}{2} \langle \hat{J}_x \rangle \langle \hat{c}_{lk \sigma}^{\dagger} \hat{d}_{\uparrow} \rangle - \frac{1}{2} \left( \lambda \langle \hat{J}_z \rangle + B_z \right) \langle \hat{c}_{lk \sigma}^{\dagger} \hat{d}_{\downarrow} \rangle + \sum_{l', \, k'} \, \gamma_{l'k'}^* \langle \hat{c}_{lk \sigma}^{\dagger} \hat{c}_{l'k' \downarrow} \rangle - (\epsilon_{l k \sigma} + \mu_l) \langle \hat{c}_{l k \sigma}^{\dagger} \hat{d}_{\downarrow} \rangle - \gamma_{l k}^* \langle \hat{d}_{\sigma}^{\dagger} \hat{d}_{\downarrow} \rangle
 \nonumber 
\end{eqnarray}
and
\begin{eqnarray}
\frac{d}{dt} \, \langle \hat{J}_x \rangle & = & -\left( \lambda \langle \hat{S}_z \rangle + B_z \right) \langle \hat{J}_y \rangle \nonumber \\ 
\frac{d}{dt} \, \langle \hat{J}_y \rangle & = & -\lambda \langle \hat{S}_x \rangle \langle \hat{J}_z\rangle + \left( \lambda \langle \hat{S}_z \rangle +  B_z \right) \langle \hat{J}_x \rangle \nonumber \\ 
\frac{d}{dt} \, \langle \hat{J}_z \rangle & = & \lambda \langle \hat{S}_x \rangle \langle \hat{J}_y \rangle
\end{eqnarray}
where we have used the choice $\lambda_x = \lambda_z = \lambda$ and $\lambda_y = 0$. Since the EOMs for the large spin components have already the desired form (see Eq.~\eqref{eoms}), hereinafter we shall focus on the time evolution of the electron operators (Eq.~\eqref{elec}). Under the Born approximation the leads are assumed to be in thermal equilibrium for all time,
\begin{eqnarray}
\langle \hat{c}_{lk \sigma}^{\dagger} \hat{c}_{l'k' \sigma'} \rangle & = & f_{l\sigma} \, \delta_{ll'} \, \delta_{\sigma \sigma'} \, \delta(k' - k),
\end{eqnarray}
with $f_{l\sigma}$ the equilibrium Fermi-Dirac distribution for spin-$\sigma$ electrons in lead $l$:
\begin{eqnarray}
f_{l\sigma} = f(\epsilon_{lk \sigma} ) =  \frac{1}{e^{(\epsilon_{lk \sigma})/k_BT} + 1}.
\end{eqnarray}
Applying the Laplace transform,
$
  \langle \hat{A} \rangle_s\equiv \int_0^{\infty} e^{-st} \langle \hat{A} \rangle_t \, dt
$,
to Eq.~\eqref{elec} we obtain:
\begin{subequations}
\begin{eqnarray}
is \, \langle \hat{d}_{\sigma}^{\dagger} \hat{d}_{\sigma'} \rangle_s & = & \frac{\lambda}{2} \left( \delta_{\sigma' \uparrow} \langle \hat{d}_{\sigma}^{\dagger} \hat{d}_{\downarrow} \rangle_s + \delta_{\sigma' \downarrow} \langle \hat{d}_{\sigma}^{\dagger} \hat{d}_{\uparrow} \rangle_s- \delta_{\sigma \uparrow} \langle \hat{d}_{\downarrow}^{\dagger} \hat{d}_{\sigma'} \rangle_s - \delta_{\sigma \downarrow} \langle \hat{d}_{\uparrow}^{\dagger} \hat{d}_{\sigma'} \rangle_s \right) \langle \hat{J}_x \rangle_s \nonumber \\ & + & \frac{1}{2} \left( \delta_{\sigma' \uparrow} \, \langle \hat{d}_{\sigma}^{\dagger} \hat{d}_{\uparrow} \rangle_s - \delta_{\sigma' \downarrow} \, \langle \hat{d}_{\sigma}^{\dagger} \hat{d}_{\downarrow} \rangle_s - \delta_{\sigma \uparrow} \, \langle \hat{d}_{\uparrow}^{\dagger} \hat{d}_{\sigma'} \rangle_s + \delta_{\sigma \downarrow} \, \langle \hat{d}_{\downarrow}^{\dagger} \hat{d}_{\sigma'} \rangle_s \right) \left( \lambda \langle \hat{J}_z \rangle_s + B_z \right) \nonumber \\ & - & \sum_{l, \, k} \, \left( \gamma_{lk} \langle \hat{c}_{lk \sigma}^{\dagger} \hat{d}_{\sigma'} \rangle_s - \gamma_{lk}^* \langle \hat{d}_{\sigma}^{\dagger} \hat{c}_{lk \sigma'}
\rangle_s \right) + i \, \langle \hat{d}_{\sigma}^{\dagger} \hat{d}_{\sigma'} \rangle_0 \label{dsdsp}
\end{eqnarray}
and
\begin{eqnarray}
is \, \langle \hat{c}_{lk \sigma}^{\dagger} \hat{d}_{\uparrow} \rangle_s & = & \frac{\lambda}{2} \langle \hat{J}_x \rangle_s \langle \hat{c}_{lk \sigma}^{\dagger} \hat{d}_{\downarrow} \rangle_s + \frac{1}{2} \left( \lambda \langle \hat{J}_z \rangle_s + B_z \right) \langle \hat{c}_{lk \sigma}^{\dagger} \hat{d}_{\uparrow} \rangle_s + f_{l\sigma} \, \delta_{\sigma \uparrow} \, \gamma_{lk}^*  - (\epsilon_{l k \sigma} + \mu_l) \langle \hat{c}_{l k \sigma}^{\dagger} \hat{d}_{\uparrow} \rangle_s - \gamma_{lk}^* \langle \hat{d}_{\sigma}^{\dagger} \hat{d}_{\uparrow} \rangle_s \label{csdu} \\
is \, \langle \hat{c}_{lk \sigma}^{\dagger} \hat{d}_{\downarrow} \rangle_s & = & \frac{\lambda}{2} \langle \hat{J}_x \rangle_s \langle \hat{c}_{lk \sigma}^{\dagger} \hat{d}_{\uparrow} \rangle_s - \frac{1}{2} \left( \lambda \langle \hat{J}_z \rangle_s + B_z \right) \langle \hat{c}_{lk \sigma}^{\dagger} \hat{d}_{\downarrow} \rangle_s + f_{l\sigma} \, \delta_{\sigma \downarrow} \, \gamma_{lk}^* - (\epsilon_{l k \sigma} + \mu_l) \langle \hat{c}_{l k \sigma}^{\dagger} \hat{d}_{\downarrow} \rangle_s - \gamma_{lk}^* \langle \hat{d}_{\sigma}^{\dagger} \hat{d}_{\downarrow} \rangle_s \label{csdd}
\end{eqnarray}
\end{subequations}
where $\langle A\rangle_0$ denotes the expectation value of operator $\hat{A}$ at time $t=0$, and where we have taken $\langle \hat{c}_{lk \sigma}^{\dagger} \hat{d}_{\sigma'} \rangle_0 = 0$. After some algebra, Eq.~\eqref{csdu} and Eq.~\eqref{csdd} become:

\begin{subequations}
\begin{eqnarray}
\langle \hat{c}_{lk \sigma}^{\dagger} \hat{d}_{\uparrow} \rangle_s & = & \gamma_{lk}^* \frac{(f_{l\sigma} \, \delta_{\sigma \uparrow} - \langle \hat{d}_{\sigma}^{\dagger} \hat{d}_{\uparrow} \rangle_s)}{\epsilon_{l k \sigma} + \mu_l - \frac{1}{2} (\lambda \langle \hat{J}_z \rangle_s + B_z) + is + \frac{1}{2} \frac{\lambda^2 \langle \hat{J}_x \rangle_s^2}{\lambda \langle \hat{J}_z \rangle_s + B_z + 2(is + \epsilon_{l k \sigma} + \mu_l)}} \nonumber \\ & + & \frac{2\gamma_{lk}^* \lambda \langle \hat{J}_x \rangle_s (\langle \hat{d}_{\sigma}^{\dagger} \hat{d}_{\downarrow} \rangle_s - f_{l\sigma})}{(\lambda \langle \hat{J}_z \rangle_s + B_z)^2 - 4 (is + \epsilon_{l k \sigma} + \mu_l)^2 + \lambda^2 \langle \hat{J}_x \rangle_s^2} \label{cdcmplt1}\\
\langle \hat{c}_{lk \sigma}^{\dagger} \hat{d}_{\downarrow} \rangle_s & = & \gamma_{lk}^* \frac{(f_{l\sigma} \, \delta_{\sigma \downarrow} - \langle \hat{d}_{\sigma}^{\dagger} \hat{d}_{\downarrow} \rangle_s)}{\epsilon_{l k \sigma} + \mu_l + \frac{1}{2} (\lambda \langle \hat{J}_z \rangle_s + B_z) + is - \frac{1}{2} \frac{\lambda^2 \langle \hat{J}_x \rangle_s^2}{\lambda \langle \hat{J}_z \rangle_s + B_z - 2(is + \epsilon_{l k \sigma} + \mu_l)}} \nonumber \\ & + & \frac{2\gamma_{lk}^* \lambda \langle \hat{J}_x \rangle_s (\langle \hat{d}_{\sigma}^{\dagger} \hat{d}_{\uparrow} \rangle_s - f_{l\sigma})}{(\lambda \langle \hat{J}_z \rangle_s + B_z)^2 - 4 (is + \epsilon_{l k \sigma} + \mu_l)^2 + \lambda^2 \langle \hat{J}_x \rangle_s^2} \label{cdcmplt2}.
\end{eqnarray}
\end{subequations}
We now consider the infinite bias limit and set, for the left lead, $\mu_L \to \infty$, and for the right, $\mu_R \to -\infty$.  In this limit, the denominator of the first term in 
Eq.~\eqref{cdcmplt1} becomes $\epsilon_{l k \sigma} + \mu_l + i 0^+ $, with positive infinitesimal $0^+$, and the second term is seen to be of the order $\mu_l^{-2}$ and thus negligiable compared with the first term (of order $\mu_l^{-1}$).  Equations \eqref{cdcmplt1} and \eqref{cdcmplt2} thus become:
\begin{eqnarray} \label{expsn}
\langle \hat{c}_{lk \sigma}^{\dagger} \hat{d}_{\sigma'} \rangle_s & = & \frac{\gamma_{lk}^*}{\epsilon_{l k \sigma} + \mu_l + i 0^+} \, (f_{l\sigma} \, \delta_{\sigma \sigma'} - \langle \hat{d}_{\sigma}^{\dagger} \hat{d}_{\sigma'} \rangle_s).
\end{eqnarray}
This result allows us to rewrite the summation that appears in Eq.~\eqref{dsdsp} as:

\begin{eqnarray}
\sum_{lk} \, \left( \gamma_{lk} \langle \hat{c}_{lk \sigma}^{\dagger} \hat{d}_{\sigma'} \rangle_s - \gamma_{lk}^* \langle \hat{d}_{\sigma}^{\dagger} \hat{c}_{lk \sigma'} \rangle_s \right) 
  & = & 
\frac{1}{2 \pi} \sum_{l}
 \int_{-\infty}^{\infty} \, d\epsilon \, \Bigg[ \frac{\Gamma_{l\sigma}(\epsilon)}{\epsilon + \mu_l + i 0^+} - \frac{\Gamma_{l\sigma'}(\epsilon)}{\epsilon+ \mu_l  - i 0^+} \Bigg] \nonumber(f(\epsilon)\delta_{\sigma \sigma'} - \langle \hat{d}_{\sigma}^{\dagger} \hat{d}_{\sigma'} \rangle_s) 
\end{eqnarray}
with the lead- and spin-dependent rates
\begin{eqnarray} \label{gamma}
\Gamma_{l\sigma} (\epsilon) & = & 2 \pi \, \rho_{l\sigma}(\epsilon) |\gamma_l(\epsilon)|^2.
\end{eqnarray}
with $\rho_{l\sigma}(\epsilon)$ density of states of the $l$-th lead. We assume these rates to be energy-independent, $\Gamma_{l\sigma}(\epsilon) = \Gamma_{l\sigma}$ (flat band approximation).  Using the Sokhatsky-Weierstrass theorem,
\begin{eqnarray}
  \frac{1}{x \pm i 0^+} & = & \mathbb{P} \, \frac{1}{x} \mp i \pi \delta(x) \nonumber,
\end{eqnarray}
upon evaluation of the Fermi functions at $\mu_L = \infty$ and  $\mu_R = -\infty$, we obtain
\begin{eqnarray}
\sum_{l, \, k} \, \left( \gamma_{lk} \langle \hat{c}_{lk \sigma}^{\dagger} \hat{d}_{\sigma'} \rangle_s - \gamma_{lk}^* \langle \hat{d}_{\sigma}^{\dagger} \hat{c}_{lk \sigma'} \rangle_s \right) & = & \frac{i}{2} \sum_l (\Gamma_{l\sigma} + \Gamma_{l\sigma'}) \langle \hat{d}_{\sigma}^{\dagger} \hat{d}_{\sigma'} \rangle_s - i \Gamma_{L\sigma} \delta_{\sigma \sigma'}.
\end{eqnarray}
Replacing the previous expression in Eq.~\eqref{dsdsp} gives:
\begin{eqnarray} \label{eomsLap}
\langle \hat{n}_{\sigma} \rangle_s & = & \lambda \langle \hat{J}_x \rangle_s \langle \hat{S}_y \rangle_s \left( \delta_{\sigma \uparrow} - \delta_{\sigma \downarrow} \right) - \Gamma \langle \hat{n}_{\sigma} \rangle_s + \Gamma_{L\sigma} \nonumber \\ 
s \, \langle S_x \rangle_s & = & -\langle \hat{S}_y \rangle_s \left( \lambda \langle \hat{J}_z \rangle_s + B_z \right) - \Gamma \, \langle \hat{S}_x \rangle_s + \langle \hat{S}_x \rangle_0 \nonumber \\
s \, \langle \hat{S}_y \rangle_s & = & -\lambda \langle \hat{S}_z \rangle_s \langle \hat{J}_x \rangle_s + \langle \hat{S}_x \rangle_s \left( \lambda \langle \hat{J}_z \rangle_s + B_z \right) - \Gamma \, \langle \hat{S}_y \rangle_s + \langle \hat{S}_y \rangle_0 \nonumber \\
s \, \langle \hat{S}_z \rangle_s & = & \lambda \langle \hat{J}_x \rangle_s \langle \hat{S}_y \rangle_s - \Gamma \langle \hat{S}_z \rangle_s + \frac{1}{2} \left( \Gamma_{L\uparrow} - \Gamma_{L\downarrow} \right)
\end{eqnarray}
where $\Gamma_{\sigma} = \Gamma_{L\sigma} + \Gamma_{R\sigma}$ for $\sigma = \uparrow, \downarrow$, although we have assumed for simplicity $\Gamma_{\uparrow} = \Gamma_{\downarrow} = \Gamma$; and the identities:
\begin{eqnarray}
\hat{S}_x & = & \frac{1}{2} \left( \hat{d}_{\uparrow}^{\dagger} \hat{d}_{\downarrow} + \hat{d}_{\downarrow}^{\dagger} \hat{d}_{\uparrow} \right) \nonumber \\ \hat{S}_y & = & \frac{1}{2i} \left( \hat{d}_{\uparrow}^{\dagger} \hat{d}_{\downarrow} - \hat{d}_{\downarrow}^{\dagger} \hat{d}_{\uparrow} \right) \nonumber \\ \hat{S}_z & = & \frac{1}{2} \left( \hat{d}_{\uparrow}^{\dagger} \hat{d}_{\uparrow} - \hat{d}_{\downarrow}^{\dagger} \hat{d}_{\downarrow} \right)
\end{eqnarray} 
have been used. Finally, inverse Laplace transforming Eqs.~\eqref{eomsLap} yields the EOMs~\eqref{eoms} for the occupation and the spin components of the electron in the QD.
 
The EOM for the total occupancy of the QD is obtained by summing the EOMs of the spin-up and spin-down occupations,
\begin{eqnarray}
\frac{d}{dt} \, \langle \hat{N} \rangle & = & -\Gamma \langle \hat{N} \rangle + \Gamma_{L\uparrow} + \Gamma_{L\downarrow}.
\end{eqnarray}
Notice that this EOM is independent of the electron and large spins, moreover, it is exactly solvable giving:
\begin{eqnarray}
\langle \hat{N} (t) \rangle & = &  \langle \hat{N} (0) \rangle \, e^{-\Gamma t} + \frac{\Gamma_{L\uparrow} + \Gamma_{L\downarrow}}{\Gamma} \left( 1 - e^{-\Gamma t} \right).
\end{eqnarray}



\subsection{Heuristic derivation}

The electronic part of our EOMs can be seen to make sense by considering a more intuitive  derivation using rate equations for the QD occupations (Eq.~\eqref{eoms}) when $\lambda = 0$.
The QD states in the transport window are $\{|0\rangle$, $|\uparrow\rangle$, $|\downarrow\rangle$, $|2\rangle = |\uparrow \downarrow\rangle \}$. In the infinite bias regime electrons can tunnel into the QD from the left lead and tunnel out of the QD to the right lead, thus
\begin{eqnarray}
\dot{p}_0 & = & W_{0\uparrow}^R p_{\uparrow} + W_{0\downarrow}^R p_{\downarrow} - (W_{\uparrow 0}^L + W_{\downarrow 0}^L) p_0 \nonumber \\
\dot{p}_{\uparrow} & = & W_{\uparrow 0}^L p_0 + W_{\uparrow 2}^R p_2 - (W_{0\uparrow}^R + W_{2 \uparrow}^L) p_{\uparrow} \nonumber \\
\dot{p}_{\downarrow} & = & W_{\downarrow 0}^L p_0 + W_{\downarrow 2}^R p_2 - (W_{0\downarrow}^R + W_{2 \downarrow}^L) p_{\downarrow} \nonumber \\
\dot{p}_2 & = & W_{2 \uparrow}^L p_{\uparrow} + W_{2 \downarrow}^L p_{\downarrow} - (W_{\uparrow 2}^R + W_{\downarrow 2}^R) p_2.
\end{eqnarray}
where $p_i$ is the probability of finding an electron in state $|i\rangle$. $W_{fi}^l$ is the tunneling rate from the initial state $|i\rangle$ to the final state $|f\rangle$ through the $l$-th barrier. Using the conservation of total probability (Tr$(\rho) = 1$) we get
\begin{eqnarray}
\dot{p}_{\uparrow} & = & W_{\uparrow 0}^L (1 - p_{\downarrow}) + (W_{\uparrow 2}^R - W_{\uparrow 0}^L) p_2 - (W_{\uparrow 0}^L + W_{0\uparrow}^R + W_{2 \uparrow}^L) p_{\uparrow} \nonumber \\
\dot{p}_{\downarrow} & = & W_{\downarrow 0}^L (1 - p_{\uparrow}) + (W_{\downarrow 2}^R - W_{\downarrow 0}^L) p_2 - (W_{\downarrow 0}^L + W_{0\downarrow}^R + W_{2 \downarrow}^L) p_{\downarrow} \nonumber \\
\dot{p}_2 & = & W_{2 \uparrow}^L p_{\uparrow} + W_{2 \downarrow}^L p_{\downarrow} - (W_{\uparrow 2}^R + W_{\downarrow 2}^R) p_2.
\end{eqnarray}
We now consider that $W_{\uparrow 0}^L = W_{2 \downarrow}^L = \Gamma_{L\uparrow}$ and $W_{\downarrow 0}^L = W_{2 \uparrow}^L = \Gamma_{L\downarrow}$, and $W_{0 \uparrow}^R = W_{\downarrow 2}^R = \Gamma_{R\uparrow}$ and $W_{0 \downarrow}^R = W_{\uparrow 2}^R = \Gamma_{R\downarrow}$, so
\begin{eqnarray}
\dot{p}_{\uparrow} & = & \Gamma_{L\uparrow} (1 - p_{\downarrow}) + (\Gamma_{R\downarrow} - \Gamma_{L\uparrow}) p_2 - (\Gamma_{L\uparrow} + \Gamma_{R\uparrow} + \Gamma_{L\downarrow}) p_{\uparrow} \nonumber \\
\dot{p}_{\downarrow} & = & \Gamma_{L\downarrow} (1 - p_{\uparrow}) + (\Gamma_{R\uparrow} - \Gamma_{L\downarrow}) p_2 - (\Gamma_{L\downarrow} + \Gamma_{R\downarrow} + \Gamma_{L\uparrow}) p_{\downarrow} \nonumber \\
\dot{p}_2 & = & \Gamma_{L\downarrow} p_{\uparrow} + \Gamma_{L\uparrow} p_{\downarrow} - (\Gamma_{R\uparrow} + \Gamma_{R\downarrow}) p_2.
\end{eqnarray}
Finally, since
\begin{equation}
\dot{p}_{\sigma} + \dot{p}_2 = \Gamma_{L\sigma} - (\Gamma_{L\sigma} + \Gamma_{R\sigma}) (p_{\sigma} + p_2),
\end{equation}
and $\langle \hat{n}_{\sigma} \rangle = p_{\sigma} + p_2$, we arrive to
\begin{equation}
\langle \dot{\hat{n}}_{\sigma} \rangle = -\Gamma \langle \hat{n}_{\sigma} \rangle + \Gamma_{L\sigma}
\end{equation}
where we have used that $\Gamma_{L\sigma} + \Gamma_{R\sigma} = \Gamma$.

\section{Isotropic model}

The EOMs for the completely isotropic case $\lambda_x = \lambda_y = \lambda_z = \lambda$ are
\begin{eqnarray} \label{eomsiso}
\frac{d}{dt} \, \langle \hat{n}_{\sigma} \rangle & = & \lambda \left( \langle \hat{J}_x \rangle \langle \hat{S}_y \rangle - \langle \hat{J}_y \rangle \langle \hat{S}_x\rangle \right) \left( \delta_{\sigma \uparrow} - \delta_{\sigma \downarrow} \right) - \Gamma \langle \hat{n}_{\sigma} \rangle + \Gamma_{L\sigma} \nonumber \\ 
\frac{d}{dt} \, \langle \hat{S}_x \rangle & = & \lambda \langle \hat{J}_y \rangle \langle \hat{S}_z\rangle - \left( \lambda \langle \hat{J}_z \rangle + B_z \right) \langle \hat{S}_y \rangle - \Gamma \langle \hat{S}_x \rangle \nonumber \\ 
\frac{d}{dt} \, \langle \hat{S}_y \rangle & = & -\lambda \langle \hat{J}_x \rangle \langle \hat{S}_z\rangle + \left( \lambda \langle \hat{J}_z \rangle +  B_z \right) \langle \hat{S}_x \rangle - \Gamma \langle \hat{S}_y \rangle \nonumber \\ 
\frac{d}{dt} \, \langle \hat{S}_z \rangle & = & \lambda \left( \langle \hat{J}_x \rangle \langle \hat{S}_y \rangle - \langle \hat{J}_y \rangle \langle \hat{S}_x\rangle \right) - \Gamma \langle \hat{S}_z \rangle + \frac{1}{2} \left( \Gamma_{L\uparrow} - \Gamma_{L\downarrow} \right) \nonumber \\
\frac{d}{dt} \, \langle \hat{J}_x \rangle & = & \lambda \langle \hat{S}_y \rangle \langle \hat{J}_z\rangle - \left( \lambda \langle \hat{S}_z \rangle + B_z \right) \langle \hat{J}_y \rangle \nonumber \\ 
\frac{d}{dt} \, \langle \hat{J}_y \rangle & = & -\lambda \langle \hat{S}_x \rangle \langle \hat{J}_z\rangle + \left( \lambda \langle \hat{S}_z \rangle +  B_z \right) \langle \hat{J}_x \rangle \nonumber \\ 
\frac{d}{dt} \, \langle \hat{J}_z \rangle & = & \lambda \left( \langle \hat{S}_x \rangle \langle \hat{J}_y \rangle - \langle \hat{S}_y \rangle \langle \hat{J}_x\rangle \right).
\end{eqnarray}
To find the solutions in the stationary limit we put to zero the time derivatives. Therefore, it can be seen right away that in the long time limit the quantum dot occupations decouple from the large spin components and becomes
\begin{equation}
\langle \hat{n}_{\sigma} \rangle = \frac{\Gamma_{L \sigma}}{\Gamma}.
\end{equation}
Thus, the spin dynamics can not be observed in the current.

\section{Effective model for region III}

In this appendix, we summarize the steps in the derivation of the effective EOMs (Eq.~\eqref{effeoms}) for region III of the parameter space (Fig.~\ref{fig2}a). Applying the transformation $\langle \mathbf{\hat{S}} \rangle = e^{-\Gamma t} \, \mathcal{R}(t) \cdot \langle \mathbf{\tilde{S}} \rangle$ and $\langle \mathbf{\hat{J}} \rangle = \mathcal{R}(t) \cdot \langle \mathbf{\tilde{J}} \rangle$ with
\begin{eqnarray}
\mathcal{R}(t) & = & \left( \begin{array}{ccc} \cos(B_zt) & -\sin(B_zt) & 0 \\ \sin(B_zt) & \cos(B_zt) & 0 \\ 0 & 0 & 1 \end{array} \right),
\end{eqnarray}
to the EOMs (Eq.~\eqref{eoms}) they become:
\begin{eqnarray}
\frac{d}{dt} \, \langle \tilde{S}_x \rangle & = & -\lambda \left\{ \langle \hat{J}_z \rangle \langle \tilde{S}_y \rangle + \left[ \langle \tilde{J}_x \rangle \cos(B_zt) - \langle \tilde{J}_y \rangle \sin(B_zt) \right] \langle \tilde{S}_z \rangle \sin(B_zt) \right\} \nonumber \\
\frac{d}{dt} \, \langle \tilde{S}_y \rangle & = & \lambda \left\{ \langle \hat{J}_z \rangle \langle \tilde{S}_x \rangle - \left[ \langle \tilde{J}_x \rangle \cos(B_zt) - \langle \tilde{J}_y \rangle \sin(B_zt) \right] \langle \tilde{S}_z \rangle \cos(B_zt) \right\} \nonumber \\ 
\frac{d}{dt} \, \langle \tilde{S}_z \rangle & = & \lambda \left[ \langle \tilde{J}_x \rangle \langle \tilde{S}_y \rangle \cos^2(B_zt) - \langle \tilde{J}_y \rangle \langle \tilde{S}_x \rangle \sin^2(B_zt) + \left( \langle \tilde{J}_x \rangle \langle \tilde{S}_x \rangle - \langle \tilde{J}_y \rangle \langle \tilde{S}_y \rangle \right) \sin(B_zt) \cos(B_zt) \right] + \frac{1}{2} (\Gamma_{L\uparrow} - \Gamma_{L\downarrow}) \, e^{\Gamma t} \nonumber \\
\frac{d}{dt} \, \langle \tilde{J}_x \rangle & = & -\lambda e^{-\Gamma t} \left\{ \langle \tilde{S}_z \rangle \langle \tilde{J}_y \rangle + \left[ \langle \tilde{S}_x \rangle \cos(B_zt) - \langle \tilde{S}_y \rangle \sin(B_zt) \right] \langle \hat{J}_z \rangle \sin(B_zt) \right\} \nonumber \\ 
\frac{d}{dt} \, \langle \tilde{J}_y \rangle & = & \lambda e^{-\Gamma t} \left\{ \langle \tilde{S}_z \rangle \langle \tilde{J}_x \rangle - \left[ \langle \tilde{S}_x \rangle \cos(B_zt) - \langle \tilde{S}_y \rangle \sin(B_zt) \right] \langle \hat{J}_z \rangle \cos(B_zt) \right\} \nonumber \\ 
\frac{d}{dt} \, \langle \tilde{J}_z \rangle & = & \lambda e^{-\Gamma t} \left[ \langle \tilde{S}_x \rangle \langle \tilde{J}_y \rangle \cos^2(B_zt) - \langle \tilde{S}_y \rangle \langle \tilde{J}_x \rangle \sin^2(B_zt) + \left( \langle \tilde{S}_x \rangle \langle \tilde{J}_x \rangle - \langle \tilde{S}_y \rangle \langle \tilde{J}_y \rangle \right) \sin(B_zt) \cos(B_zt) \right]
\end{eqnarray}
Since in the long-time limit $d \, \langle \tilde{J}_i \rangle/dt \to 0$, we assume $\langle \mathbf{\tilde{J}} \rangle$ to be stationary. Therefore, the EOMs for the electron spin in the original frame become:
\begin{eqnarray}
\frac{d}{dt} \, \langle \mathbf{\hat{S}} \rangle & = & \mathbf{B}_{\rm eff} \times \langle \mathbf{\hat{S}} \rangle - \Gamma \langle \mathbf{\hat{S}} \rangle + \frac{1}{2} (\Gamma_{L\uparrow} - \Gamma_{L\downarrow}) \, \mathbf{u}_z
\end{eqnarray}
where $\mathbf{B}_{\rm eff} = (B_{\rm ac}(t),0,B_z)$ and $\mathbf{u}_z$ is the unit vector pointing in the $z$-direction.

\vspace{1cm}

\end{widetext}

\bibliography{mybib_Berlin}{}

\end{document}